\documentclass[twocolumn,english,prc,nofootinbib,floatfix,showpacs]{revtex4-1}
\usepackage[T1]{fontenc}
\usepackage[utf8]{luainputenc}
\usepackage{color}
\usepackage{array}
\usepackage{amsmath}
\usepackage{amssymb}
\usepackage{graphicx}

\makeatletter

\providecommand{\tabularnewline}{\\}

\usepackage{graphics}\usepackage{epsfig}\usepackage{amsfonts}%\usepackage{dcolumn}
\usepackage{bm}% bold math
\newcommand{\pr}[1]{{\sc{\lowercase{#1}}}}

\renewcommand{\vec}[1]{\bm{#1}}

\usepackage{babel}

\makeatother

\usepackage{babel}
\begin{document}

\title{The effect of realistic nuclear charge distributions on isotope shifts and towards the extraction of higher order nuclear radial moments}

\author{A. Papoulia$^{1}$, B.~G.~Carlsson$^{1}$}

\email{gillis.carlsson@matfys.lth.se}

\selectlanguage{english}%

\author{J.~Ekman$^{2}$}

\affiliation{$^{1}$Division of Mathematical Physics, LTH, Lund University, Post
Office Box 118, S-22100 Lund, Sweden}

\affiliation{$^{2}$Group for Materials Science and Applied Mathematics, Malmö
University, S-20506 Malmö, Sweden}

\date{\today}
\begin{abstract}
\begin{description}\item[Background] Atomic spectral lines from different
isotopes display a small shift in energy, commonly referred to as
the line isotope shift. One of the components of the isotope shift
is the field shift, which depends on the extent and the shape of the
nuclear charge density distribution. \item[Purpose] To investigate
how sensitive field shifts are with respect to variations in the nuclear
size and shape and what information of nuclear charge distributions
that can be extracted from measured field shifts. \item[Methods]
Nuclear properties are obtained from nuclear density functional theory calculations based
on the Skyrme-Hartree-Fock-Bogoliubov approach. These results are
combined with multiconfiguration Dirac-Hartree-Fock methods to obtain
realistic field shifts. \item[Results] Phenomena such as nuclear
deformation and variations in the diffuseness of nuclear charge distributions
give measurable contributions to the field shifts. Using a novel approach,
we demonstrate the possibility to extract new information concerning
the nuclear charge densities from the observed field shifts. \item[Conclusions]
Combining methods used in atomic and nuclear structure theory gives an improved description
of field shifts. Extracting additional nuclear information from measured
field shifts is possible in the near future with improved experimental methods.
\end{description}
\end{abstract}


\maketitle

\section{Introduction}

Information of nuclear sizes has grown rapidly during the last decades.
In the compilation by Angeli and Marinova in 2013 \cite{Angeli},
root mean square (rms) radii were reported for more than 900 isotopes
of which the majority are radioactive systems. This development is
a consequence of refined experimental and theoretical methods, and
a state of the art example is the frequency comb measurement of the
Hydrogen-Deuterium radius difference by Parthey et al. \cite{Hydr_Deut}.
The plenitude of available data has allowed for detailed investigations
of the evolution of nuclear radii for isotope sequences along virtually
the entire periodic table. These studies have revealed unexpected
trends, especially close to magic numbers, which serves as benchmarks
for nuclear structure calculations \cite{Agbemava2014}.

However, more detailed and model independent experimental information
of nuclear charge distributions beyond the rms radius is only available
for stable or long-lived isotopes from electron scattering experiments. On the theoretical
side it has been shown that isotope shifts in heavier systems depend
on the nuclear model used \cite{Andrae} and that the contribution
from nuclear deformation to the isotope shift in some cases are comparable
to the uncertainty in recent dielectronic recombination experiments
\cite{Dielectr_exp,Zubova}.

Experimental techniques such as high-precision laser measurements at the COLLAPS and CRIS
experiments at ISOLDE/CERN \cite{ISOLDE} and dielectronic recombination
experiments at the envisaged realization of CRYRING at GSI \cite{GSI} are constantly
evolving. This justifies a more systematic theoretical investigation of what information
can be revealed about nuclear charge distributions in exotic systems.

The main objective of the present work is to study the effect of realistic
charge distributions, taken from nuclear density functional theory
(DFT), on the isotope shift in heavier atoms. To the knowledge of the
authors this has never been done before. In addition, a promising
and novel method for the extraction of higher order radial moments
from experimental isotope shifts is also presented and tested.

\section{Isotope shifts}

The atomic nucleus is $\sim10^{4}$ smaller than the size of the atom.
Even so, the finite mass and extended charge distribution of the nucleus
have a measurable effect on atomic spectra. Spectral lines from different
isotopes display a small shift in energy referred to as the isotope
shift (IS), which can further be decomposed into a mass shift (MS)
and a field shift (FS) contribution. The difference in energy between
the corresponding atomic level $i$ of two isotopes $A$ and $A'$,
the level isotope shift, can thus be expressed as 
\begin{equation}
\delta E_{i,IS}^{A,A'}=\delta E_{i,MS}^{A,A'}+\delta E_{i,FS}^{A,A'}=E_{i}^{A'}-E_{i}^{A}.\label{eq:is1}
\end{equation}
For a particular atomic transition $k$ between upper $u$ and lower
$l$ levels, the difference in energy for a pair of isotopes, namely
the line frequency isotope shift, is consequently given by

\begin{equation}
\begin{split}\delta\nu_{k,IS}^{A,A'}=\delta\nu_{k,MS}^{A,A'}+\delta\nu_{k,FS}^{A,A'}=\nu_{k}^{A'}-\nu_{k}^{A}\\
=\frac{\delta E_{u,IS}^{A,A'}-\delta E_{l,IS}^{A,A'}}{h}.
\end{split}
\label{eq:fs1}
\end{equation}

The level mass shift contribution can be expressed as 
\begin{equation}
\delta E_{i,MS}^{A,A'}=\left(\frac{M'-M}{MM'}\right)K_{MS}^{i},
\end{equation}
where $M$ and $M'$ are the atomic masses of the isotopes and $K_{MS}^{i}$
is the mass-independent mass shift parameter~\cite{Mass_shift_ref3,nuclear_recoil_corr,Nuclear_recoil_corr_ref4}.
Although the computation of the mass shift parameters, and hence the
mass shift contribution to the isotope shift, represents a challenging
task, it is not the main focus of this work. Instead, the focus here
is on the extent and shape of nuclear charge distributions which almost
exclusively effect the field shift described in detail below.

\subsection{Field shift }

The field shift arises from differences in the nuclear charge density
distribution between isotopes caused by the different number of neutrons.
Unlike point-like charge distributions, more realistic charge distributions
alter the central field that the atomic electrons experience, and
hence the atomic level and transition energies will be affected. Evidently,
the field shift effect is more pronounced for electrons moving in
$s_{1/2}$ and $p_{1/2}$ orbitals due to the non-zero probability
of the radial wave functions at the origin. Moreover, the nuclear
charge and extent, together with the contraction of the atomic orbitals,
increase with the proton number $Z$ and thus the contribution from
the field shift to the isotope shift is found to be dramatically larger
in heavier systems.

\subsubsection{Non-pertubative ``exact'' method}

In atomic structure calculations, where the contribution from the
mass shift is neglected, the level field shift can be computed according
to Eq.~\ref{eq:is1} by performing separate calculations for two
isotopes, $A$ and $A'$, with different parameter sets describing
the respective nuclear charge distributions. This method is in general
highly model-dependent, since the description of the nucleus is normally
restricted to an approximate model. Moreover, this procedure is cumbersome
if calculations are to be performed for many isotope pairs and in
addition it may suffer from numerical instabilities since it involves
the substraction of large quantities (atomic binding energies) to
obtain a tiny quantity. Nevertheless, this strategy constitutes an
``exact'' method for estimating the validity of perturbative approaches and
the resulting field shifts will be denoted $\delta\nu_{k,VA}^{exact}$ below.

\subsubsection{Perturbative method}

To eliminate the disadvantages of the exact method described above
and allow for a more flexible analysis of the field shift, an alternative
approach based on perturbation theory may be used. Within the framework
of pertubation, the first-order level field shift of level $i$ can
be written 
\begin{equation}
\delta E_{i,FS}^{(1)A,A'}=-\int_{R^{3}}[V_{A'}(\vec{r})-V_{A}(\vec{r})]\rho_{i}^{e}(\vec{r})d^{3}\vec{r},\label{eq:fspert}
\end{equation}
where $V_{A}(\vec{r})$ and $V_{A'}(\vec{r})$ are the one-electron
potentials arising from the different nuclear charge distributions
of the two isotopes and $\rho_{i}^{e}(\vec{r})$ is the electron density
inside the nuclear volume of the reference isotope $A$.

Following the work by Seltzer~\cite{Isotope_shifts}, Torbohm et
al.~\cite{e_density} and Blundell et al.~\cite{e_density_2} and
assuming an extended spherical symmetric nuclear charge distribution,
it can be shown that the electron density to a very good approximation
can be expanded around $r=0$ as an even polynomial function keeping
only the first few terms: 
\begin{equation}
\rho_{i}^{e}(\vec{r})\approx b_{i}(r)=b_{i,1}+b_{i,2}r^{2}+b_{i,3}r^{4}+b_{i,4}r^{6}.\label{eq:polyexp}
\end{equation}
Inserting the expression above in Eq.~\ref{eq:fspert} and making
use of the Laplacian operator in spherical coordinates, $\nabla^{2}r^{2N}=2N(2N+1)r^{2N-2}$,
Poisson's equation, $\nabla^{2}V_{A}(\vec{r})=-4\pi\rho_{A}(\vec{r})$,
and finally Eq.~\ref{eq:fs1}, the first-order line frequency field
shift is given by \cite{e_density_2,tobepublished} 
\begin{equation}
  \delta\nu_{k,FS}^{(1)A,A'}\approx\delta\nu_{k,RFS}^{A,A'}=\sum_{N=1}^{4}F_{k,N}\delta\left\langle r^{2N}\right\rangle ^{A,A'},\label{eq:linefs}
\end{equation}
where $F_{k,N}$ are the so-called line electronic factors expressed
as 
\begin{equation}
F_{k,N}=\frac{2\pi}{h}\frac{Z\Delta b_{k,N}}{N(2N+1)},\label{eq:ef}
\end{equation}
and 
\begin{equation}
\delta\left\langle r^{2N}\right\rangle ^{A,A'}=\left\langle r^{2N}\right\rangle ^{A}-\left\langle r^{2N}\right\rangle ^{A'}
\end{equation}
are the differences of the nuclear radial moments, of order $2N$,
of the isotopes $A$ and $A'$. The electronic factors are proportional
to the difference of the electronic density inside the nucleus between
the upper and lower atomic level, thus $\Delta b_{k,N}=b_{u,N}-b_{l,N}$.

The reformulated field shift (RFS) according to Eq.~\ref{eq:linefs}
enables a more versatile analysis of field shifts. This is due to
the fact that the radial moments, $\langle r^{2N}\rangle$, used in
the expression can be taken from any model, calculation or experiment.
In addition, it is possible to analyze the contributions to the field
shift order by order. For example, keeping only the first term in
Eq.~\ref{eq:linefs} we obtain 
\begin{equation}
\delta\nu_{i,FS}^{(1)A,A'}\approx\frac{2\pi}{3h}Z\Delta\rho_{i}^{e}(\vec{0})\delta\left\langle r^{2}\right\rangle ^{A,A'},\label{eq:fspert1}
\end{equation}
which is a suitable approximation for lighter systems where a constant
electron density within the nucleus can be assumed, $\rho_{i}^{e}(\vec{r})\approx b_{i,1}=\rho_{i}^{e}(\vec{0})$.
For heavier systems, however, the electron density varies inside the
nuclear volume and thus the $N\geq2$ terms in Eq.~\ref{eq:linefs}
must also be considered for an accurate description. Further on, by
including these higher-order contributions the effect on the isotope
shift due to details in the nuclear charge distribution can be analyzed.
As we shall see, the reversed approach is also possible, namely to
extract higher order radial moments of the nuclear charge distribution
from observed isotope shifts.

\subsection{Computational procedure}

Solutions to the many-body Hamiltonian describing the atom are obtained
by performing calculations using the relativistic atomic structure
package \pr{GRASP2K}~\cite{Grasp2K_2013}, which is based on the
multiconfiguration Dirac-Hartree-Fock (MCDHF) approach. In the MCDHF
method, atomic state functions, $\Psi(\gamma PJM_{J})$, which are
approximate solutions to the Dirac-Coulomb Hamiltonian, are expanded
over configuration state functions (CSFs), $\Phi(\gamma_{i}PJM_{J})$,
with appropriate total angular momentum ($J$) symmetry and parity
$P$: 
\begin{equation}
\Psi(\gamma PJM_{J})={\displaystyle \sum_{i=1}^{N}c_{i}\Phi(\gamma_{i}PJM_{J})}.
\end{equation}
In the expression above, $\gamma_{i}$ represents the configuration,
coupling and other quantum number necessary to uniquely describe the
state $i$, $M_{J}$ is the projection of $J$ on the $z$-axis and
$c_{i}$ are mixing coefficients fulfilling the condition $\sum_{i=1}^{N}c_{i}^{2}=1$.
The CSFs are constructed from one-electron Dirac orbitals that together
with the mixing coefficents are obtained in a relativistic self-consistent-field
procedure by applying the variational principle \cite{Relat_atoms}.
The transverse photon interaction as well as leading quantum electrodynamic
(QED) corrections can be accounted for in subsequent relativistic
configuration interaction (RCI) calculations \cite{Breit_QED}.

Once a set of ASFs is obtained, the computation of the isotope shift
parameters is carried out using the program \pr{RIS4} \cite{tobepublished},
which represents an extension of the predecessor \pr{RIS3} \cite{ris3u}.
In \pr{RIS4} the polynomial expansion $b_{i}(r)$ given by Eq.~\ref{eq:polyexp}
is for each level fitted to the constructed electron density $\rho_{i}^{e}(\vec{r})$
using a least-squares method. Finally, by combining the expansion
coefficients $b_{i,N}$ from two or more levels the line electronic
factors are computed for the reference isotope $A$ according to Eq.~\ref{eq:ef}.

\section{realistic nuclear charge distributions}

As seen above, the reformulated field shift depends on the radial
moments of the nuclear charge distribution. These moments can be calculated
from nuclear models that provide accurate charge distributions. In
this section three such models are compared.

\subsection{Nuclear charge distribution models}

The nuclear charge distribution can be approximated by an analytical
expression such as the Fermi distribution: 
\begin{equation}
\rho(r,\theta)=\frac{\rho_{0}}{1+e^{\frac{r-c(\theta)}{a}}},
\end{equation}
where, if only axially symmetric quadrupole deformation is considered:
$c(\theta)=c_{0}[1+\beta_{20}Y_{20}(\theta)]$. The value of $\rho_{0}\approx\rho(r=0)$
is determined by the normalization condition: 
\begin{equation}
\int\rho(\vec{r})d\vec{r}=1.
\end{equation}
and the parameter $\alpha$ is given by the relation: 
\begin{equation}
t=4\ln(3)\alpha,
\end{equation}
where $t$ is the skin thickness of the distribution. The skin thickness
is defined as the interval where the density decreases from 90\% to
10\% of $\rho(0)$. The parameter $c_{0}$ reflects the size of the
nucleus.

In the \pr{GRASP2K} code \cite{Grasp2K_2013} the explicit values
for these parameters are taken as \cite{Andrae}: $t=2.3$ fm, $\beta_{20}=0$
and the parameter $c_{0}$ is chosen so that the rms radius of the
nuclear charge distribution becomes 
\begin{equation}\label{eq:grasp_formula}
\sqrt{\left\langle r^{2}\right\rangle }=0.836\cdot A^{\frac{1}{3}}+0.570\,\mathrm{\, fm}\,\,(A>9),
\end{equation}
where $A$ denotes the number of nucleons of the isotope.\medskip{}

Realistic nuclear charge distributions can also be obtained from microscopic
nuclear models based on effective interactions. Such models have the
advantage that the size, shape and diffuseness of the nuclear density
is obtained by solving a self-consistent set of Hartree-Fock-Bogoliubov
(HFB) equations.

In this work we adopt the effective Skyrme interaction \cite{Ring_Schuck}
and consider two different sets of Skyrme parameters called SLY4 and
UNEDF1. The parameters in both sets are adjusted to fit experimental
data in a broad range of nuclei. The SLY4 set was fitted with an emphasis
on describing neutron rich nuclei \cite{SLY4}, whereas the UNEDF1
set constitutes a more recent parametrization fitted to reproduce
both ground state energies as well as radii and single-particle energies
\cite{UNEDF1}. In spherical symmetry, the solutions to the HFB equations
are provided by the code \pr{HOSPHE} (v2.00), which is a new version
of the program \pr{HOSPHE} (v1.02) \cite{Carlsson_hosphe_v1.02}.
In the case of deformed nuclei, we use the code \pr{HFBTHO} (2.00d)
\cite{HFBTHO}, based on a cylindrically deformed harmonic oscillator
(HO) basis.

For spherical nuclei we take into account the finite nature
of protons by folding the densities using the convolution formula
\begin{equation}
\varrho_{c}(\vec{r})=\int d^{3}\vec{r'}\rho_{p}(\vec{r'})g(\left|\vec{r}-\vec{r'}\right|),
\end{equation}
where $\rho_{p}(\vec{r})$ is the initially calculated proton density
and 
\begin{equation}
g(\vec{r})=(r_{0}\sqrt{\pi})^{-3}e^{-(\vec{r}/r_{0})^{2}},
\end{equation}
the proton form factor, assumed to be a Gaussian with $r_{0}=\sqrt{\frac{2}{3}}\cdot r_{p}^{rms}$,
where $r_{p}^{rms}$ is the proton rms radius \cite{Folding}. Experiments
to determine the proton radius have resulted in different values of
$r_{p}^{rms}$ \cite{proton_radius0.88,proton_radius} and in this
work we adopt the results based on electron scattering measurements
assuming $r_{p}^{rms}=0.88$ fm.

\begin{figure}
\includegraphics[width=0.9\columnwidth]{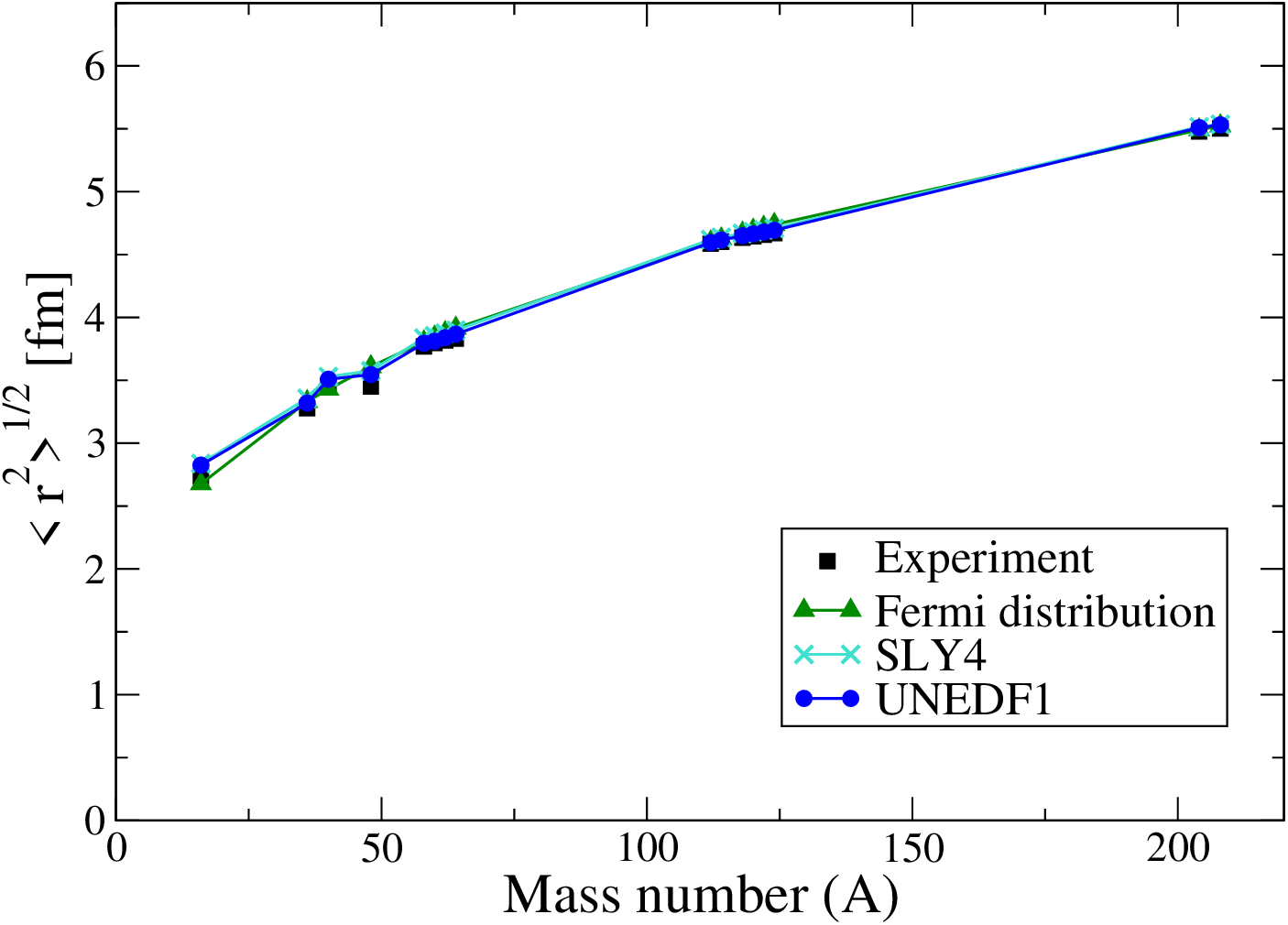}\caption{{\small{{}Rms radii of theoretical charge distributions compared
to experimental data. Two different Skyrme parameter sets, SLY4 and
UNEDF1, are used with moments calculated after taking into account
the finite proton size. The resulting $\sqrt{\left\langle r^{2}\right\rangle }$
values from the Fermi distribution used in the }}\pr{{\small{{}GRASP2K}}}
code{\small{{} (Eq.~ \ref{eq:grasp_formula}) are also included.\label{fig:Rms-radii}}}}
\end{figure}

In Fig.~\ref{fig:Rms-radii} the theoretical rms radii are compared
to experimental data obtained from elastic electron scattering experiments
\cite{Exper_data}. A total of 16 spherical isotopes of various elements:
O, S, Ca, Ni, Sn and Pb, are used in the comparison. As seen in this
figure both the nuclear models as well as the empirical parametrization
{\small{{}(Eq.~\ref{eq:grasp_formula}) }}are in good agreement
with the experimental data.{\small{ }}{\small \par}

\begin{figure}
\includegraphics[width=1\columnwidth]{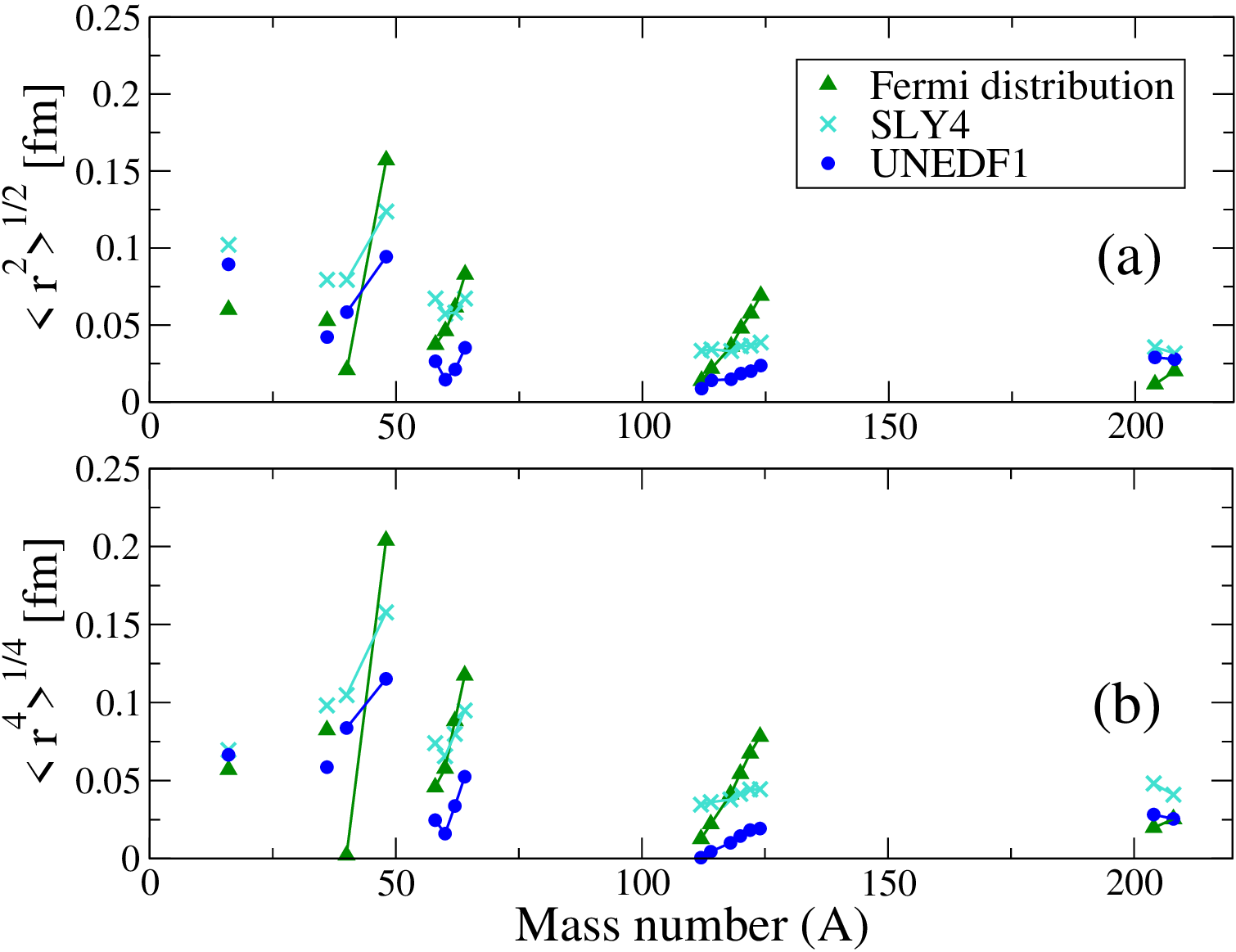}

\caption{{\small{{}Discrepancy of (a) the rms radii $\sqrt{\left\langle r^{2}\right\rangle }$
and (b) the $\sqrt[4]{\left\langle r^{4}\right\rangle }$ moment of
the theoretical charge distributions compared to experimental data.
Isotopic sequences are connected with lines. }}}
\end{figure}

The discrepancy between the theoretical and the experimental $\sqrt{\left\langle r^{2}\right\rangle }$
values is shown in Fig.~2a. As seen in this figure the more recent
Skyrme parameters (UNEDF1) give the best description of the data.
The two microscopic models also stand out as they are in general better
at capturing the isotopic trends giving flatter curves than the Fermi
distribution.

For calculations of field shifts, the higher order moments may also
play an important role and in Fig.~2b the discrepancy in the prediction
of the $\sqrt[4]{\left\langle r^{4}\right\rangle }$ values is shown.
This comparison shows the same trend as for the $\sqrt{\left\langle r^{2}\right\rangle }$
values, namely that microscopic models capture the isotopic trends
better while the Fermi distribution in general does a good job for
the stable nuclei. One might consider using more refined empirical
expressions containing a dependence on the difference in proton and
neutron numbers, but since such an approach would anyway not capture
the important changes caused by deformations the best approach comes
from using state-of-the-art microscopic nuclear models.

\begin{table}
\begin{tabular}{>{\raggedright}p{0.36\columnwidth}>{\raggedright}p{0.28\columnwidth}>{\raggedright}p{0.12\columnwidth}}
\hline 
\noalign{\vskip0.08cm}
 & $\sqrt{\left\langle r^{2}\right\rangle }$  & $\sqrt[4]{\left\langle r^{4}\right\rangle }$\tabularnewline[0.08cm]
\hline 
\hline 
\noalign{\vskip0.08cm}
Fermi distribution  & $0.01660$  & $0.01954$\tabularnewline
Skyrme-SLY4  & $0.01821$  & $0.01905$\tabularnewline
Skyrme-UNEDF1  & $0.01271$  & $0.01260$\tabularnewline
\hline 
\end{tabular}\caption{{\small{{}\label{tab:The-standard-deviations}Standard deviations
of discrepancies in $\sqrt{\left\langle r^{2}\right\rangle }$ and
$\sqrt[4]{\left\langle r^{4}\right\rangle }$, calculated for the
three theoretical models. }}}
\end{table}

In Tab.~\ref{tab:The-standard-deviations}, the {\small{{}}}standard
deviations of the discrepancies for the three models are compared.
Considering the average agreement, the Fermi distribution and the
Skyrme-SLY4 give similar results while the more recent UNEDF1 is significantly
better. In addition, the UNEDF1 set predicts the $\sqrt{\left\langle r^{2}\right\rangle }$
and $\sqrt[4]{\left\langle r^{4}\right\rangle }$ moments with about
the same precision, while the precision deteriorates slightly for
the two other models. This agrees with the fact that the full density
profiles also tend to be better reproduced by UNEDF1. Higher order
moments are difficult to compare since more focus is then shifted
towards the surface and tail of the density where insufficient precision
in the data hampers a qualitative comparison. All in all, the UNEDF1
parametrization describes the nuclear charge distributions more accurately
than both the Skyrme-SLY4 and Fermi distributions and therefore realistic
nuclear radial moments resulting from this interaction will be used
in the following in order to estimate the line field shifts.

\subsection{Application to {\small{{}line field shifts}}}

In this section the atomic physics calculations for the electron energies
are combined with the use of the microscopic nuclear models for the
charge densities. As an example we consider the resonance transition
$6s^{2}$ $^{1}S_{0}\longrightarrow$ $6s6p$ $^{1}P_{1}^{\circ}$ observed
in several neutral Ba isotopes. By comparing the line field shift
in the isotope series one may be able to draw conclusions on the shape
and size of the nuclear density distributions.
The most abundant barium isotope on Earth, $^{138}$Ba, is taken as
a reference and the shifts in electron energies are thus compared
to the values for this isotope. This reference isotope is spherical,
while the other isotopes obtained by removing or adding a couple of
neutrons are predicted to have more deformed shapes.

\begin{figure}
\includegraphics[width=0.9\columnwidth]{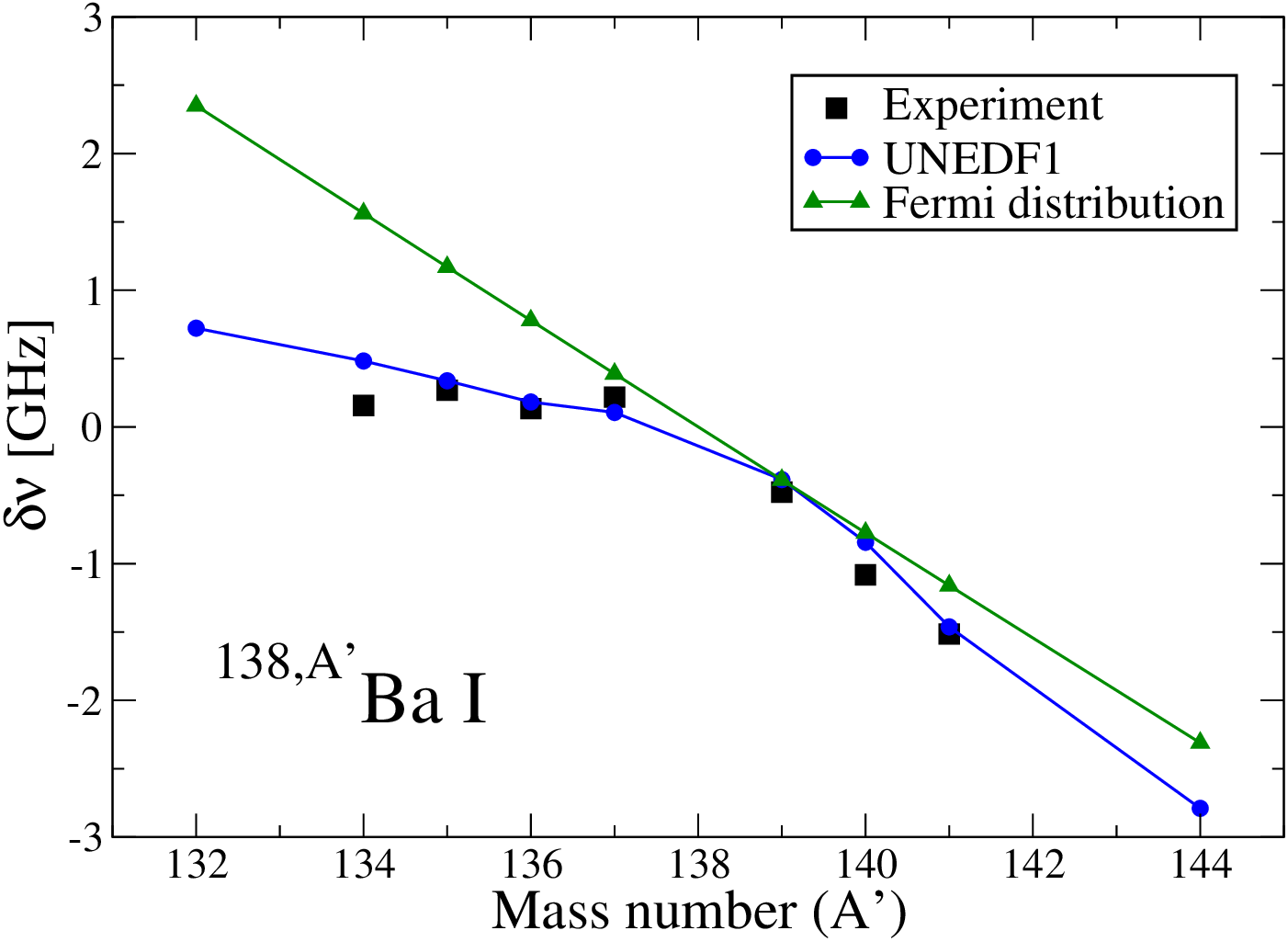}\caption{{\small{{}\label{fig:The-absolute-line}The absolute line field shift
values are compared to the available experimental data \cite{Ba_IS_measurements,Barium_Theoret_IS}.
Nuclear radial moments resulted from the realistic HFB calculations
using the Skyrme-UNEDF1 interaction, as well as from the Fermi distribution,
have been used. All plotted values refer to the $6s^{2}$ $^{1}S_{0}\longrightarrow$
$6s6p$ $^{1}P_{1}^{\circ}$ transition.}}}
\end{figure}

Fig.~\ref{fig:The-absolute-line} shows the calculated line field
shifts for the Ba isotope series compared to experimental isotope
shifts \cite{Ba_IS_measurements}, where theoretical mass shift contributions
have been subtracted \cite{Barium_Theoret_IS}. The calculations based
on the Fermi distribution shows a linear dependence on the mass number
$A'$ of the target isotope and fail to capture the general trend.
The microscopic nuclear calculations capture both the right trend
with neutron number and in addition some of the odd-even staggering.

\section{Effect of realistic charge distributions on the line field shifts}

In order to investigate the resulting field shifts when replacing
the commonly adopted Fermi distribution with more realistic nuclear
models we examine the differences in the predicted field shifts for
a variety of isotopes. For such analysis the Fermi distribution is
fitted so that it has the same $\left\langle r^{2}\right\rangle $
value as computed from the realistic distributions. Then 
\begin{align}
\delta\nu_{Fermi} & =F_{k,1}\delta\left\langle r^{2}\right\rangle _{realistic}\nonumber \\
+ & {\displaystyle \sum_{N=2}^{4}}F_{k,N}\delta\left\langle r^{2N}\right\rangle _{Fermi}.
\end{align}
Thus the correction when using realistic charge distributions is given
by
\begin{align}
\delta\nu_{realistic}-\delta\nu_{Fermi} & ={\displaystyle }\nonumber \\
\sum_{N=2}^{4}F_{k,N}[\,\delta\left\langle r^{2N}\right\rangle _{realistic} & {\displaystyle -\delta\left\langle r^{2N}\right\rangle _{Fermi}\,]}.
\end{align}
In the following two subsections the size of this correction term
will be investigated for lithium-like and neutral systems.

\subsection{Li-like systems }

Isotope shifts in lithium-like systems have been studied theoretically
and experimentally in the past \cite{Dielectr_exp,Dielectr_exp_2,Zubova,li2012mass,Exp_IS_ND57+,Elliott} and are
thus of particular interest. In Fig.~4 the magnitude of the ``correction
term'' $\delta\nu_{realistic}-\delta\nu_{Fermi}$ for one of the
resonance transitions has been plotted as a function of the mass number
$A'$ of the target isotope for a wide range of Li-like systems. For
the spherical Sn, Pb, Er and Uuh nuclear systems the magnitude of
the corrections increases with $A'$. Moreover, the absolute magnitude
of the $\delta\nu_{realistic}-\delta\nu_{Fermi}$ term increases with
the difference between the neutron number $\Delta N^{A,A'}$ in the
isotope sequences of Sn and Pb. When more neutrons are added they
alter the protons distribution, leading to changes in the diffuseness.
This effect is not included in the Fermi model where a constant skin
thickness $t\simeq2.3$ fm is assumed and may be a reason for the
observed difference. 

\begin{figure}
\includegraphics[width=1\columnwidth]{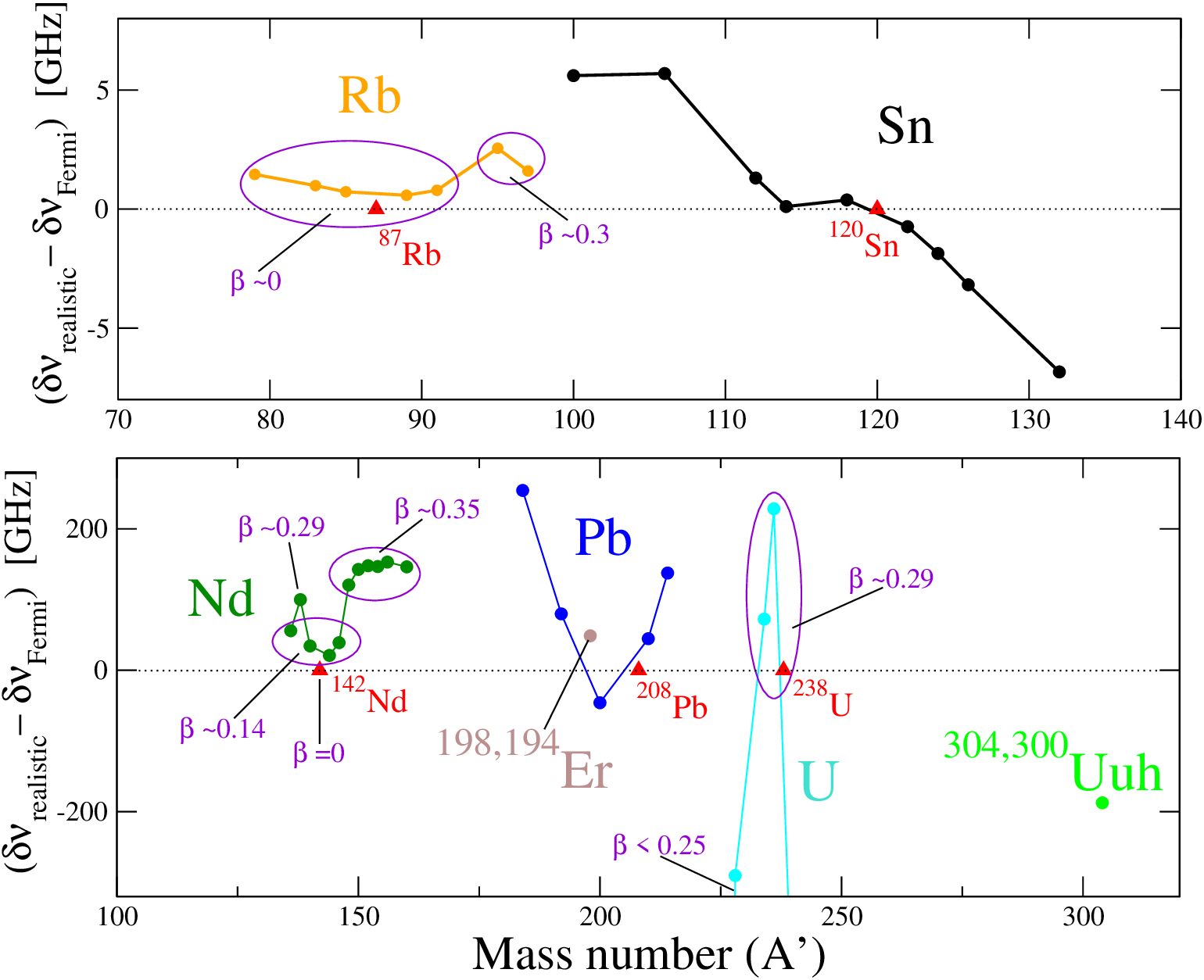}\caption{{\small{The corrections $\delta\nu_{realistic}-\delta\nu_{Fermi}$
to the line field shift calculations as a function of the mass number
$A'$ of the target isotope for various Li-like systems. For the systems
that contain deformed isotopes, the magnitude of the quadrupole deformation
parameter $\beta_{20}$ of the target isotopes $A'$ is indicatively
shown. The isotopes used as reference are marked with triangles 
and all plotted values refer to the $1s^{2}2s$ $^{2}S_{1/2}$$\longrightarrow$
$1s^{2}2p$ $^{2}P_{1/2}^{\circ}$ resonance transition. }}}
\end{figure}

In the deformed Rb, Nd and U systems the corrections depend on the
size of the nuclei as well as the quadrupole deformation parameter
$\beta_{20}$, which is assumed to be zero in the spherical Fermi
model. Hence, for large deformations the corrections for the Rb and
Nd isotope pairs are comparable to the ones obtained for the spherical
Sn and Pb isotope pairs. For the heavier U isotopes the corrections
become significantly large in spite of the small difference in deformation
between the  reference and target isotopes. 

In Fig.~5 the magnitude of the corrections has been plotted as a
function of the calculated deformation parameter $\beta_{20}$ corresponding
to the isotope $A'$ for some Nd and U isotope pairs. In both plots
the magnitude of the ``correction term'' increases as the difference
between the deformation of reference and target isotope becomes large.
The largest corrections are obtained for the uranium isotope
pairs $^{240,238}$U and $^{220,238}$U. In this case the correction
amounts to $\simeq2.3$\% and $2$\%, respectively.

\begin{figure}
\includegraphics[width=1\columnwidth]{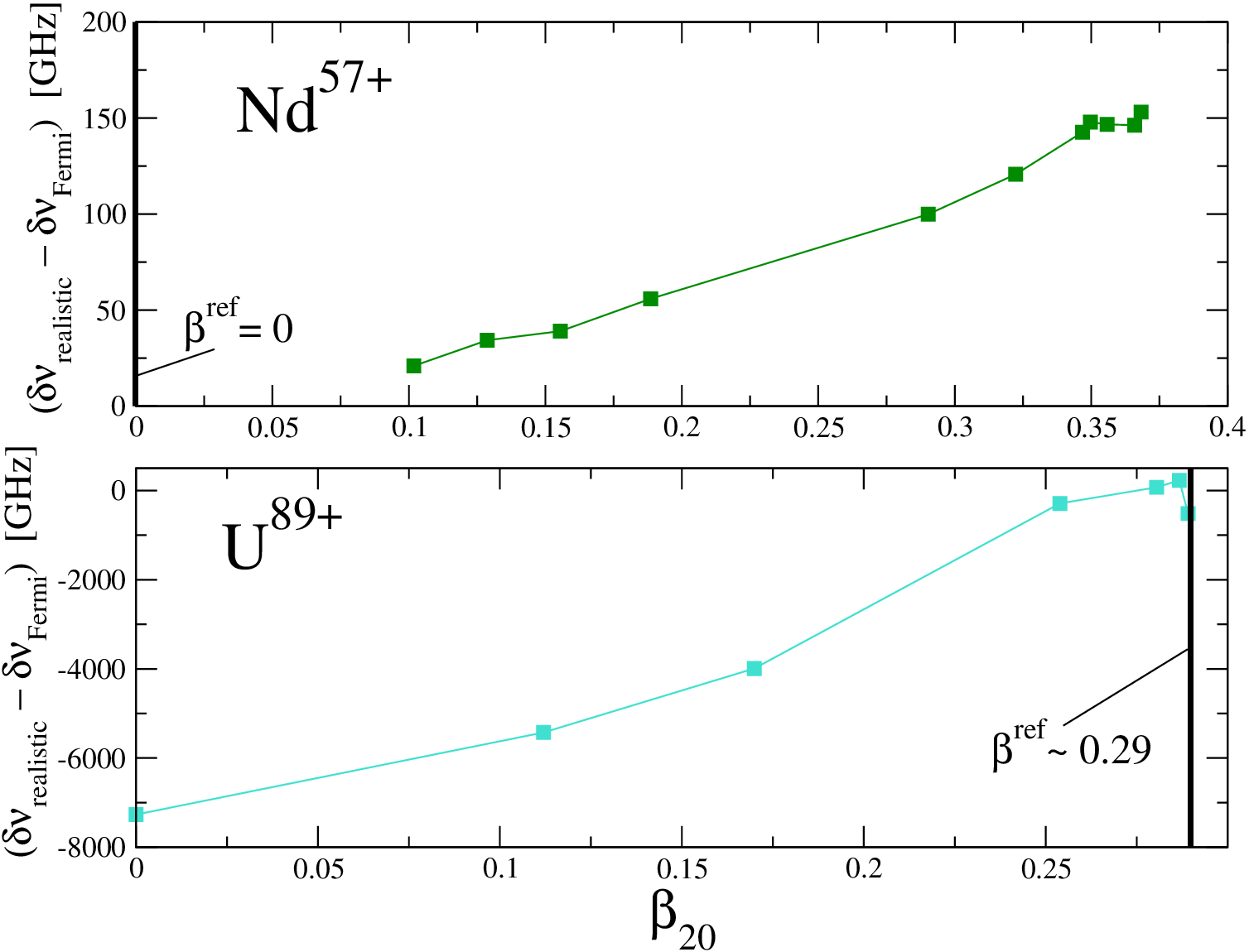}

\caption{{\small{{}The corrections $\delta\nu_{realistic}-\delta\nu_{Fermi}$
to the line field shift values as a function of the quadrupole deformation
parameter $\beta_{20}$ of the target $A'$ isotope for various (a)
Nd$^{57+}$ and (b) U$^{89+}$ isotope pairs. In each case, the corresponding
deformation of the reference isotope $A$ is indicated by a vertical
line on the plots. All plotted values refer to the same resonance
transition as in Fig.~4.}}}
\end{figure}

The two-parameter Fermi model does not take into account the effect
of deformation. As a result, the effect of realistic charge distributions
on the field shifts is larger in atomic systems with deformed nuclei.
The correction term $\delta\nu_{realistic}-\delta\nu_{Fermi}$ can
however be decomposed into two parts and written as 
\begin{align}
\delta\nu_{realistic}-\delta\nu_{Fermi} & =\left(\delta\nu_{realistic}-\delta\nu_{Fermi}^{def}\right)+\nonumber \\
+ & \left(\delta\nu_{Fermi}^{def}-\delta\nu_{Fermi}^{sph}\right).
\end{align}
The $\delta\nu_{Fermi}^{def}-\delta\nu_{Fermi}^{sph}$ part isolates
the effect of deformation, while the remaining $\delta\nu_{realistic}-\delta\nu_{Fermi}^{def}$
part gives the corrections due to ``other effects'', such as density
wiggles and differences in diffuseness. In order to separately estimate
the effect of deformation in Li-like Nd, the deformed Fermi model
was used with $\beta_{20}$ values obtained from the microscopic nuclear
calculations. 

\begin{figure}
\includegraphics[width=0.9\columnwidth]{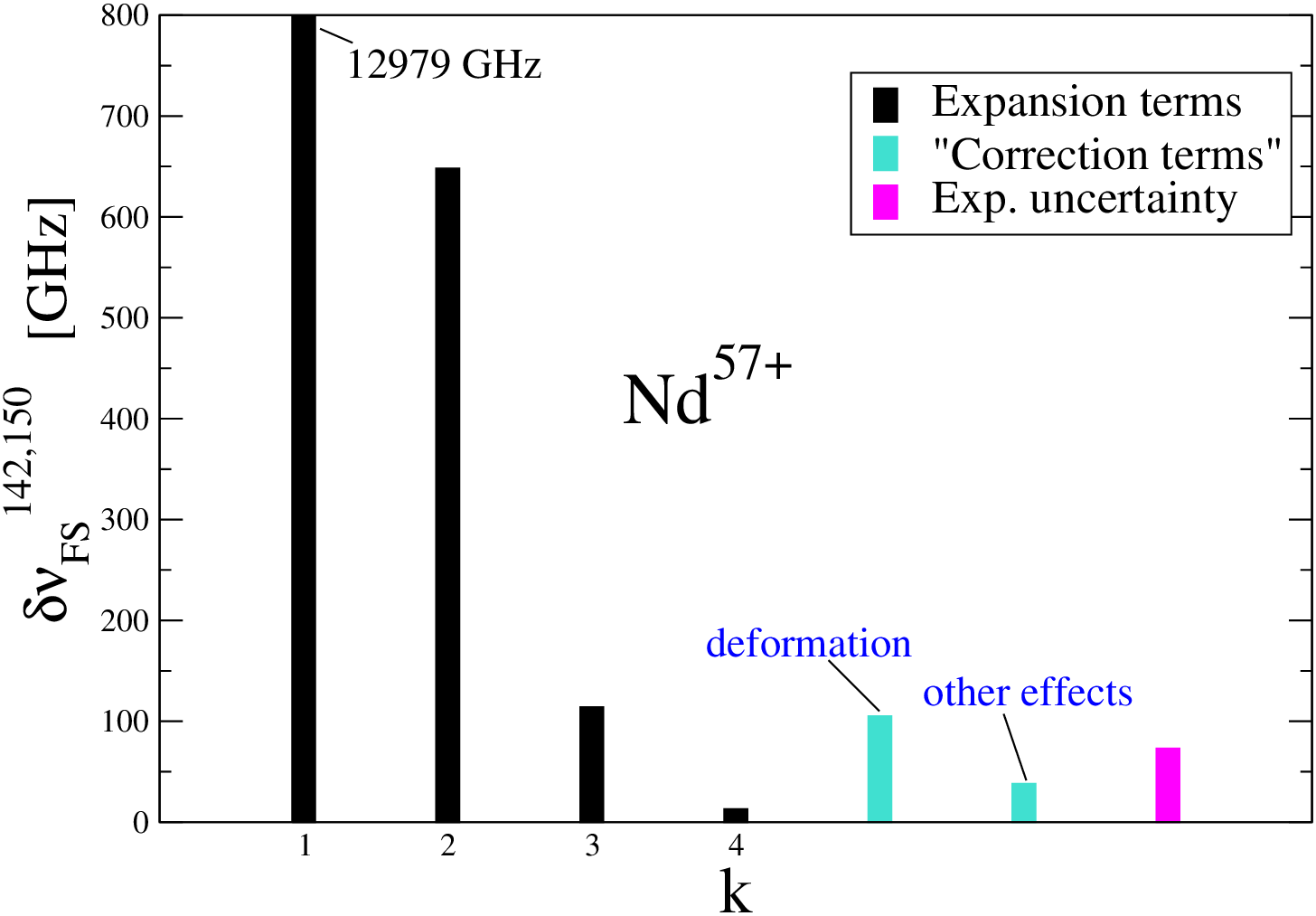}

\caption{{\small{{}Decomposition of expansion and correction terms of the
$1s^{2}2s$ $^{2}S_{1/2}$$\longrightarrow$ $1s^{2}2p$ $^{2}P_{1/2}^{\circ}$
transition in Li-like $^{142,150}$Nd.}}}
\end{figure}

Isotope shift (IS) measurements have been performed for the first two
resonance transitions of the $^{142,150}$Nd$^{57+}$pair \cite{Exp_IS_ND57+} and
the statistical uncertainty of the observed isotope shift for the $1s^{2}2s$ $^{2}S_{1/2}$$\longrightarrow$ $1s^{2}2p$
$^{2}P_{1/2}^{\circ}$ transition is compared
to the magnitude of the ``correction terms'' in Fig.~6. As seen
in the figure, the effect of deformation is large enough to be detected
by the experiments and the correction due to ``other effects'' is not neglible.

\subsection{Neutral atoms}

In this subsection, field shifts in neutral barium are investigated
for the three well-known $6s^{2}$ $^{1}S_0$ $\longrightarrow$
$6s6p$ $^{1,3}P_1^\circ$ and $6s^{2}$ $^{1}S_0$ $\longrightarrow$ $6p^{2}$
$^{3}P_1$ transitions. Fig.~7 illustrates the dependence of the magnitude of the
corrections on the deformation parameter $\beta_{20}$. The same trend
is seen for the three transitions. As already deduced for
Nd$^{57+}$ and U$^{89+}$ (see Fig.~5) the magnitude of $\delta\nu_{realistic}-\delta\nu_{Fermi}$
increases as the difference between the deformation of reference and
target isotope becomes large. However, in neutral barium the magnitude
of the ``correction term'' $\delta\nu_{realistic}-\delta\nu_{Fermi}$
is a factor $\sim10^{3}$ smaller.

\begin{figure}
\includegraphics[width=0.9\columnwidth]{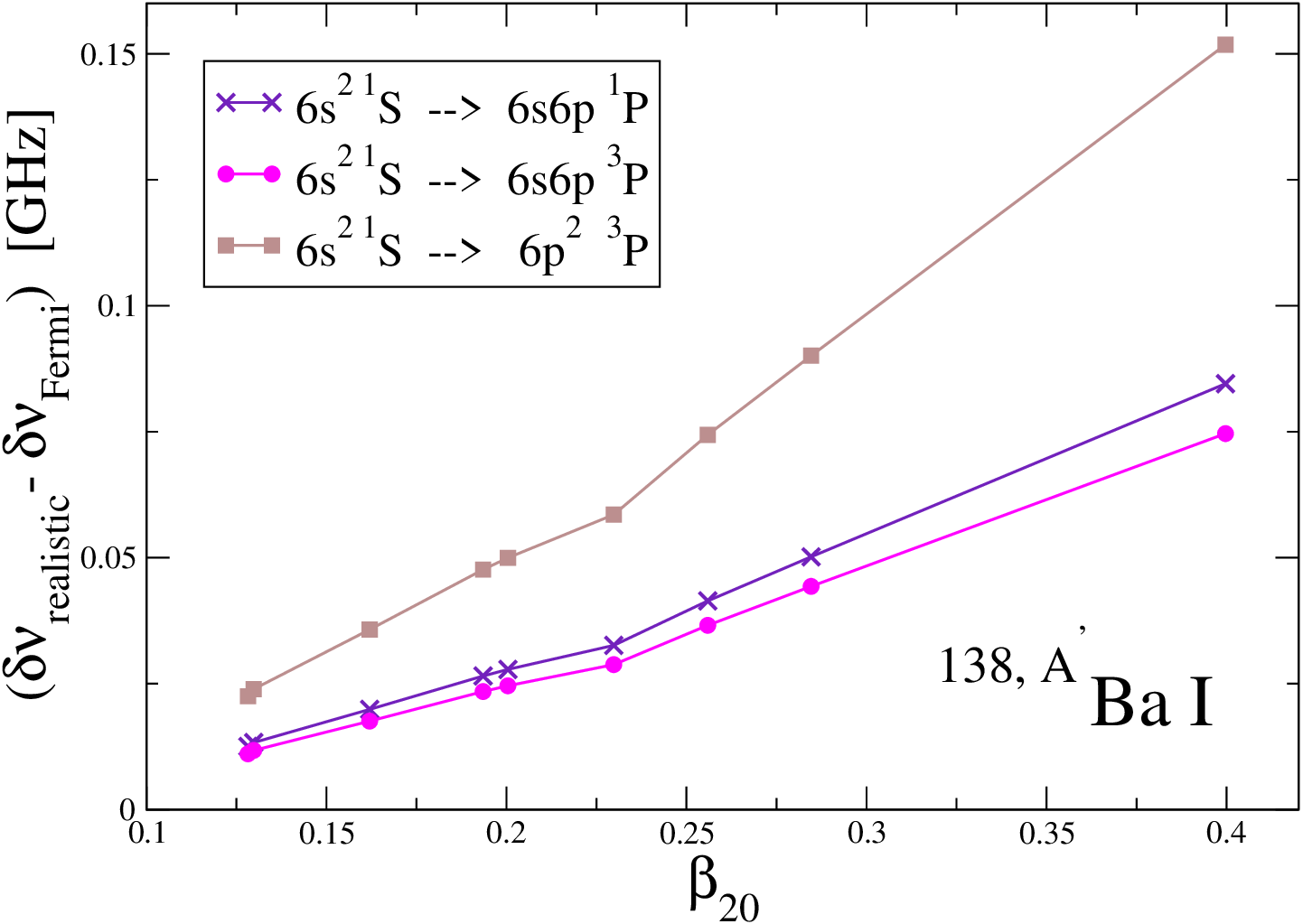}\caption{{\small{{}The corrections, $\delta\nu_{realistic}-\delta\nu_{Fermi}$,
to the line field shift calculations as a function of the quadrupole
deformation parameter $\beta_{20}$ of the target $A'$ isotope for
the neutral barium isotope pairs.}}}
\end{figure}

In contrast to the IS measurements in Li-like systems, a greater number
of measurements has been performed in neutral atomic systems. Furthermore,
in such measurements the accuracy provided is generally much higher. Following
the process described in the previous section the ``correction term''
is decomposed for the $6s^{2}$ $^{1}S_0$ $\longrightarrow$ $6s6p$
$^{1}P^\circ_1$ transition of the $^{138,136}$Ba isotope pair. The isotope
shift measurements of the corresponding spectral lines \cite{Ba_IS_measurements}
carries a statistical error, which is in Fig.~8 compared to the magnitude
of the ``correction terms''. 

As seen in Fig.~8 the experimental uncertainty is remarkably small
in comparison to the magnitude of the corrections. However, in reality
the experimental uncertainty of the field shift is much larger since the theoretical
mass shift contribution is in this case associated with large uncertainties,
which are not reflected in this figure. The dominating corrections
are the ``other effects'' that arise from the differences between
the deformed Fermi distribution and the more realistic charge distributions
obtained from the microscopic nuclear calculations.

\begin{figure}
\includegraphics[width=0.9\columnwidth]{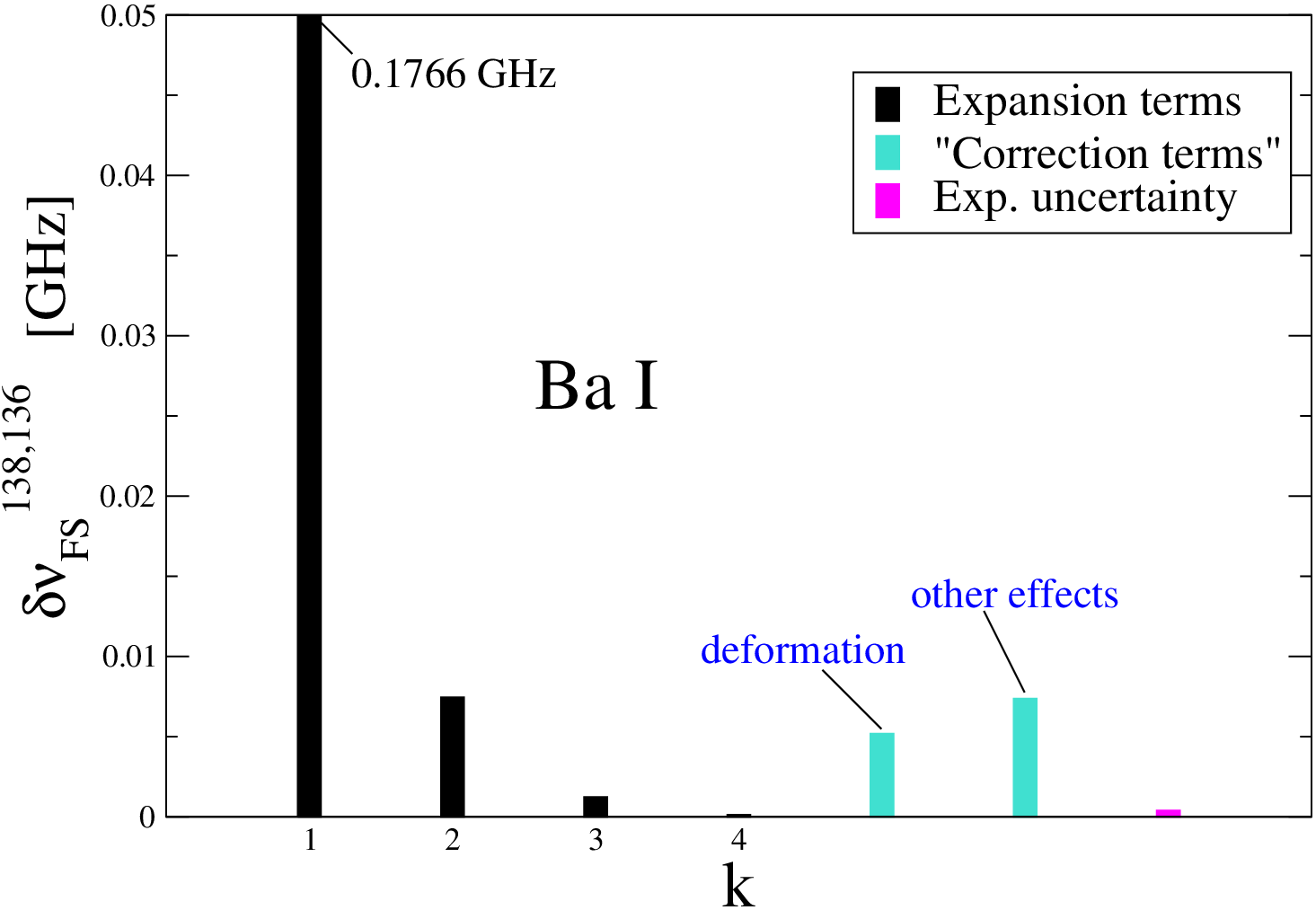}\caption{{\small{{}Decomposition of expansion and correction terms of the
$6s^{2}$ $^{1}S_0$ $\longrightarrow$ $6s6p$ $^{1}P^\circ_1$ transition
in $^{138,136}$Ba I.}}}
\end{figure}
\medskip{}

The major improvement to the line field shift measurements illustrated
in Fig.~3 is clearly due to the choice of using realistic rms radii.
However, making in addition use of realistic higher order nuclear
moments leads to a non-negligible improvement in the description of
the experimental data. According to the current experimental precision
in the measurement of the isotope shifts in $^{136,138}$Ba and $^{150,142}$Nd$^{57+}$,
effects like deformation captured by the higher nuclear moments could
be detected (see Fig.~6 and 8). As a result, information about such
a nuclear property could possibly be deduced from isotope shift observations.

\section{$\delta\left\langle r^{4}\right\rangle ^{A,A'}$ extraction}

The nuclear charge radius is one of the most obvious and fundamental
parameters, related to the size of the nucleus. Considering isotope
shift measurements the charge radii of an isotope sequence are typically
determined in terms of the differences in the second radial moment,
$\delta\left\langle r^{2}\right\rangle $, between target isotope
$A'$ and reference isotope $A$. In contrast to light nuclei, in
heavy nuclear systems the contribution of the higher order radial
moments to the line field shift can be significant and above the observable
limit (see Fig.~6 and 8). Moreover, in highly charged heavy systems
the contribution of the mass shift effect becomes smaller. This suggests
the possibility to extract information about higher nuclear moments. 

The reformulation of the field shift, combined with experimental isotope
shift measurements, in principle enables the extraction of differences
in higher order radial moments $\delta\left\langle r^{2N}\right\rangle $,
$N=2,3,4$. Consequently information about the nuclear shapes, deformations,
density wiggles and other nuclear properties can be provided. The
extraction of all four radial moments requires four transitions $k$
to be available. A system of four equations is then solved for 
\begin{align}\label{eq:foureq}
\delta\nu_{k,RFS} & = F_{k,1}\delta\left\langle r^{2}\right\rangle +F_{k,2}\delta\left\langle r^{4}\right\rangle +\nonumber \\
+ & F_{k,3}\delta\left\langle r^{6}\right\rangle +F_{k,4}\delta\left\langle r^{8}\right\rangle ,
\end{align}
where $k=1,2,3,4$. However, it is rare that observed isotope shifts
are available for four transitions and in addition, such systems of
equations cannot be formed so that they give trustworthy solutions
for higher than second order moments.

\subsection{RFS expansion using orthogonal moments}

As seen in Fig.~6 and 8, all four expansion terms do not equally
contribute to the final field shift value. Considering in Fig.~9
the line field shift for the $^{208,200}$Pb pair, the $4th$ order
radial moment adds $\sim10\%$ contribution, the $6th$ moment $\sim2\%$
and the last term, which contains the $8th$ order moment, contributes
with much less. Thus it is fair to say that the major correction to
the approximation that assumes constant electron density $\rho_{i}^{e}(\vec{r})\approx\rho_{i}^{e}(0)$
comes from the second expansion term, i.e. $F_{k,2}\delta\left\langle r^{4}\right\rangle $,
which takes into account the differences between the $\left\langle r^{4}\right\rangle $
moments. However, the contribution from higher order terms is not
negligible.

\begin{figure}
\includegraphics[width=0.85\columnwidth]{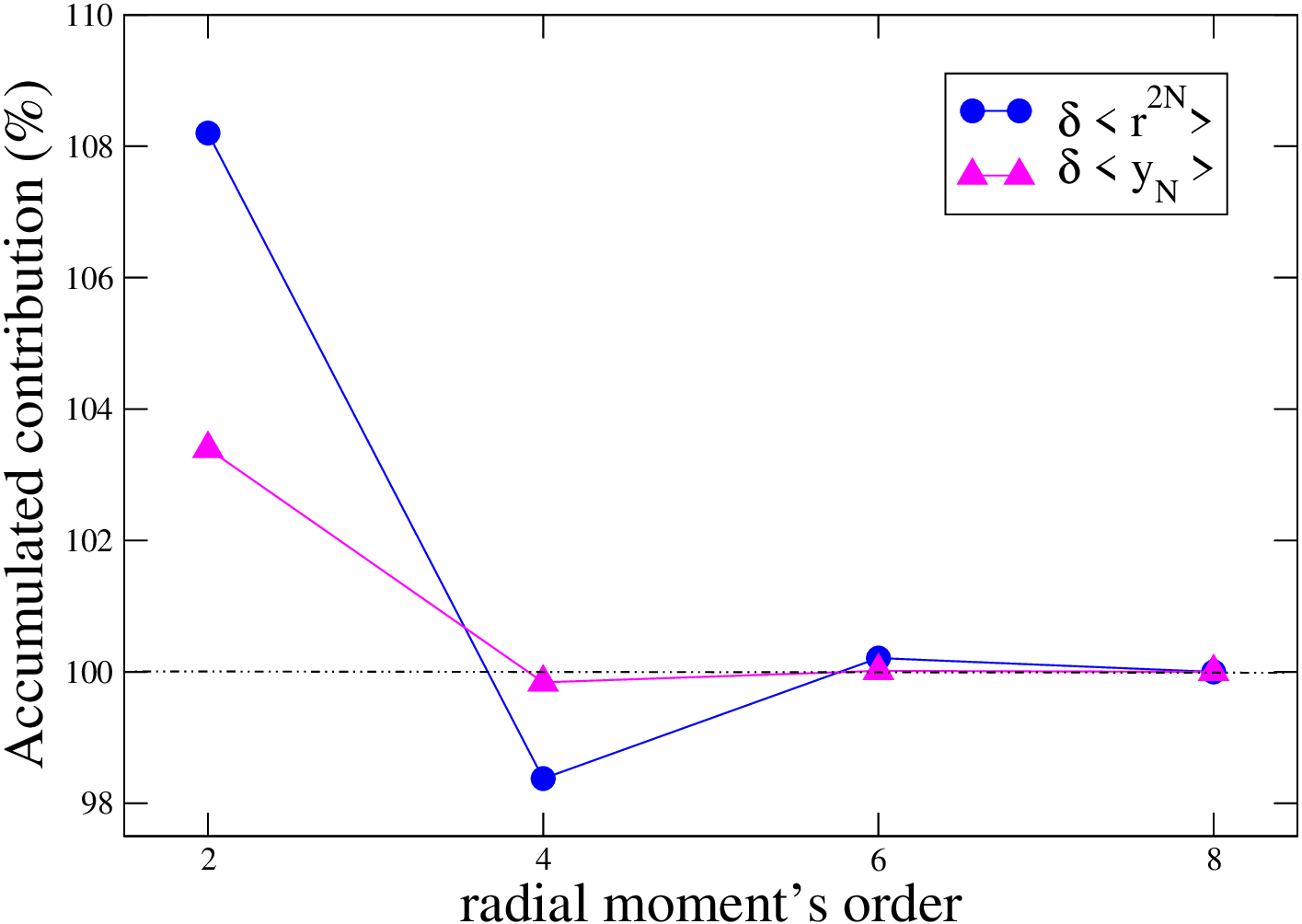}

\caption{{\small{{} ${\displaystyle {\textstyle \sum_{N}}}F_{N}\delta\left\langle r^{2N}\right\rangle /\delta\nu$
in percent (circles) compared to the corresponding expression for
the re-arranged summation (triangles). The plot refers to the $^{208,200}$Pb
pair and the $1s^{2}2s$ $^{2}S_{1/2}$$\longrightarrow$ $1s^{2}2p$
$^{2}P_{1/2}^{\circ}$ transition. }}}
\end{figure}

In Eq.~\ref{eq:foureq} the information about the nuclear charge distribution is
encoded in a set of nuclear radial moments. These moments are not
independent and a faster converging series may be found by instead
expanding in a set of orthogonal polynomials (see Appendix A). The
convergence of this re-arranged summation compared with the original
summation is shown in Fig.~9. By taking into account only the first
term, the line field shift is already much closer to the final value.
The second term adds $\sim3.5\%$ contribution, the third $\sim0.18\%$,
while the last one adds $\sim0.016\%$. Thus accurate enough field
shift predictions can now be provided using only the first two expansion
terms containing the differences $\delta\left\langle y_{1}\right\rangle $
and $\delta\left\langle y_{2}\right\rangle $, which are in turn given
as a function of the $\delta\left\langle r^{2}\right\rangle $ and
$\delta\left\langle r^{4}\right\rangle $ moments (see Appendix A).
Having only two unknowns means that $\delta\left\langle r^{2}\right\rangle $
and $\delta\left\langle r^{4}\right\rangle $ can potentially be extracted
from knowledge of two observed line field shifts in an isotope pair.

\subsection{Testing the new method}

After expanding in the orthonormal basis, for a pair of isotopes $A,A'$, the reformulated line field shift can to a very good approximation
be expressed as 
\begin{equation}\label{eq:yeq}
\delta\nu_{k,RFS}\approx c_{k,1}\delta\left\langle y_{1}\right\rangle +c_{k,2}\delta\left\langle y_{2}\right\rangle ,
\end{equation}
where the $c_{k,1}$ and $c_{k,2}$ coefficients are expressed in
terms of the $F_{k,N}$ factors. In order to test the new method,
theoretical line field shifts, $\delta\nu_{RFS}$, were obtained using
realistic nuclear radial moments. These line field shifts refer
to the $1s^{2}2s$ $^{2}S_{1/2}$$\longrightarrow$ $1s^{2}2p$ $^{2}P_{1/2}^{\circ}$
and $1s^{2}2s$ $^{2}S_{1/2}$$\longrightarrow$ $1s^{2}2p$ $^{2}P_{3/2}^{\circ}$
transitions, of the uranium, lead and neodymium isotope pairs studied
in Sec. IV. Using these calculated field shifts as ``pseudo-experimental'' input data,
the equations can be inverted and should yield, if the method is flawless, extracted
radial moments which are identical to the realistic nuclear moments used in the computation of
the field shifts.

\begin{figure}
\includegraphics[width=1\columnwidth]{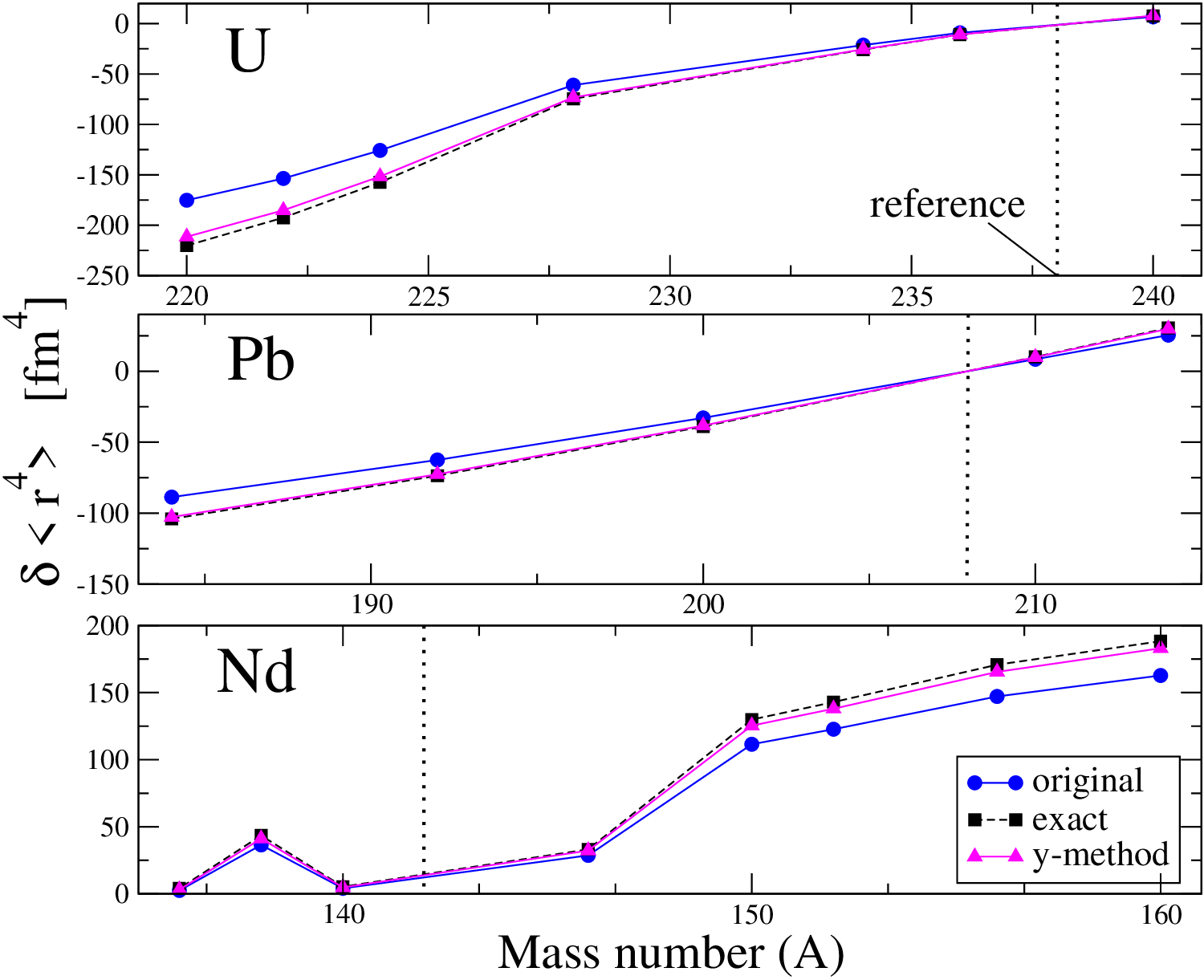}

\caption{{\small{{}The extracted $\delta\left\langle r^{4}\right\rangle $
values for the uranium, lead and neodymium isotope pairs, previously
studied. The dashed line with squares represents the exact $\delta\left\langle r^{4}\right\rangle _{realistic}$
values, obtained from the HFB and HFBTHO calculations for the spherical
lead and the deformed uranium and neodymium isotopes, respectively.
The line with triangle symbols represents the extracted $\delta\left\langle r^{4}\right\rangle$
when $\delta\nu_{k,FS}\approx c_{k,1}\delta\left\langle y_{1}\right\rangle +c_{k,2}\delta\left\langle y_{2}\right\rangle $
is assumed, while the line with circles corresponds to the $\delta\left\langle r^{4}\right\rangle _{original}$
extracted when the original summation in $\delta\nu_{k,RFS}$ has been
used.}}}
\end{figure}

In all cases the extracted $\delta\left\langle r^{2}\right\rangle $
moments are almost identical to the exact $\delta\left\langle r^{2}\right\rangle _{realistic}$
moments. The difference is less than $0.0002$ fm$^{2}$ for all lead
and uranium isotopes, as well as the neodymium isotopes that are close
to spherical. For the highly deformed neodymium isotopes, the difference
is slightly larger, of the order of $\simeq0.001$ fm$^{2}$, which
still represents a small discrepancy. 

In Fig.~10, the extracted $\delta\left\langle r^{4}\right\rangle $
values have been plotted and compared to the $\delta\left\langle r^{4}\right\rangle _{realistic}$
representing exact values. The extracted $\delta\left\langle r^{4}\right\rangle _{original}$
values using the first two terms of the original summation $\delta\nu_{k,RFS}^{A,A'}\approx{\displaystyle {\textstyle \sum_{N=1}^{2}}}F_{k,N}\delta\left\langle r^{2N}\right\rangle $
are in addition illustrated in the same figure. When the re-arranged
summation is used the extracted $\delta\left\langle r^{4}\right\rangle $
moments are in good agreement with the exact $\delta\left\langle r^{4}\right\rangle _{realistic}$
moments, whereas the $\delta\left\langle r^{4}\right\rangle $ moments
using the original, but truncated, summation display an observable
discrepancy from the exact values. All in all, the new expression
using the re-arranged summation for the reformulated field shift enables
the determination of the differences between $r^{2}$ and $r^{4}$
moments, much more accurately than using the original expression.

\subsection{Towards the extraction of $\delta\left\langle r^{2}\right\rangle $
and $\delta\left\langle r^{4}\right\rangle $ moments using experimental
data}

In what follows, the major objective is to discuss how $\delta\left\langle r^{2}\right\rangle $
and $\delta\left\langle r^{4}\right\rangle $ moments can be extracted
from experimental data using the new method tested above. From observed isotope shifts,
experimental field shift values can be obtained by estimating and removing the mass shift contribution
and residual effects, $\delta\nu_{k,RES}$, from example QED and nuclear polarization~\cite{Zubova}:
\begin{equation}
 \delta\nu_{k,FS}^{exp} = \delta\nu_{k,IS}^{exp}-\delta\nu_{k,MS}-\delta\nu_{k,RES}.
%\delta\nu_{k,IS}^{exp}-\delta\nu_{k,MS}\approx c_{k,1}\delta\left\langle y_{1}\right\rangle +c_{k,2}\delta\left\langle y_{2}\right\rangle ,
\end{equation}
Without making use of RFS, the difference in $\langle r^2\rangle$ moments can now be extracted by peforming variational calculations where the
rms radius of the reference isotope is estimated and $\delta\langle r^2\rangle$ is varied until agreement with
experimental field shifts is observed (see for example~\cite{Exp_IS_ND57+}):
\begin{equation}
 \delta\nu_{k,FS}^{exp} = \delta\nu_{k,VA}^{exact}.
\end{equation}
The difference in higher moments then follows from the model used to mimic the nuclear charge distribution, for example
the Fermi distribution, and hence this method is highly model dependent.
However, making use of the reformulation of field shifts using an orthogonal moments basis, we instead use experimental field shift values from
two transitions and solve the following equation system in order to extract the $\delta\left\langle r^{2}\right\rangle$ and
$\delta\left\langle r^{4}\right\rangle $ moments virtually model independent:
\begin{equation}\label{eq:theequation}
  \delta\nu_{k,FS}^{exp} = c_{k,1}\delta\left\langle y_{1}\right\rangle +c_{k,2}\delta\left\langle y_{2}\right\rangle + d_k.
%\delta\nu_{k,IS}^{exp}-\delta\nu_{k,MS}\
\end{equation}
In the expression above a term $d_k$ has been introduced which represents
the discrepancy between the ``exact'' variational solution, $\delta\nu_{k,VA}^{exact}$, and the RFS solution, $\delta\nu_{k,RFS}$,
assuming a spherical Fermi nuclear charge distribution for the reference and the target isotope.
To examine the importance of the $d_k$-term for the extraction of the radial moments, we used
\pr{GRASP2K} and \pr{RIS4} to compute $\delta\nu_{k,VA}^{exact}$ and $\delta\nu_{k,RFS}$ for the resonance transitions
in several Li-like lead and uranium isotope pairs. In the calculations rms radii were taken from the compilation by Angelis \& Marinova \cite{Angeli} and the results are presented in Tab.~II and III. As seen, an expected discrepancy
between the $\delta\nu_{VA}^{exact}$ and $\delta\nu_{RFS}$ values, i.e. the $d_k$ term,
is observed for both transitions. In our case, this discrepancy is mainly due to the quantum electrodynamic
(QED) effects included in the VA calculation that become important in heavy nuclei and which are not
included in the perturbative approach. Other assumptions that have
been made throughout the formulation of the perturbative approach
are expected to play a minor role. Indicatively, for ``transition
1'' in the uranium isotope pairs the magnitude of the discrepancy
is of the order of $\sim1.5$\% of the $\delta\nu_{VA}^{exact}$ value,
from which $\sim0.1$\% is due to other than QED effects. It is also seen that the $d_k$ terms for the two transitions are
slightly different, and it turns out that accurately estimating this difference, rather than the magnitude of the terms, is absolutely crucial in order to extract accurate $\delta\left\langle r^{4}\right\rangle $ moments.

\begin{table}
\begin{tabular}{>{\raggedright}p{1.8cm}lll}
\hline 
\noalign{\vskip0.2cm}
$\mathrm{Transition1}$ & $\mathbf{\delta\nu_{VA}^{exact}}$$\mathbf{[GHz]}$  & $\mathbf{\mathbf{\delta\nu_{RFS}}}$$\mathbf{[GHz]}$  & $\mathbf{d}$$\mathbf{[GHz]}$ \tabularnewline
\hline 
\hline
\noalign{\vskip0.1cm}
$\mathbf{208,192}$  & $51303$  & $50563$  & $740$\tabularnewline
$\mathbf{208,200}$  & $28938$  & $28546$  & $392$\tabularnewline
$\mathbf{208,210}$  & -$14186$  & -$14021$  & -$165$\tabularnewline
\noalign{\vskip0.1cm}
$\mathbf{238,234}$  & $54796$  & $53976$  & $820$\tabularnewline
$\mathbf{238,236}$  & $27412$  & $27015$  & $397$\tabularnewline
\hline 
\end{tabular}\caption{{\small{{}The line frequency field shift values, resulting from the
variational calculations using \pr{GRASP2K} and the reformulation of the
field shift, are respectively displayed for a few lead and uranium
isotope pair combinations. In the last column, the discrepancy between
$\delta\nu_{VA}^{exact}$ and $\delta\nu_{RFS}$ is computed. ``Transition
1'' refers to the $1s^{2}2s$ $^{2}S_{1/2}$$\longrightarrow$ $1s^{2}2p$
$^{2}P_{1/2}^{\circ}$ transition.}}}
\end{table}

\begin{table}
\begin{tabular}{>{\raggedright}p{1.8cm}lll}
\hline 
\noalign{\vskip0.2cm}
Transition2 & $\mathbf{\delta\nu_{VA}^{exact}}$$\mathbf{\mathbf{[GHz]}}$  & $\mathbf{\delta\nu_{RFS}}$$\mathbf{[GHz]}$  & $\mathbf{d}$$\mathbf{[GHz]}$ \tabularnewline
\hline 
\hline 
\noalign{\vskip0.1cm}
$\mathbf{208,192}$  & $55459$  & $54642$  & $817$\tabularnewline
$\mathbf{208,200}$  & $31282$  & $30848$  & $434$\tabularnewline
$\mathbf{208,210}$  & -$15336$  & -$15152$  & -$184$\tabularnewline
\noalign{\vskip0.1cm}
$\mathbf{238,234}$  & $61189$  & $60277$  & $912$\tabularnewline
$\mathbf{238,236}$  & $30610$  & $30169$  & $441$\tabularnewline
\hline 
\end{tabular}\caption{{\small{{}Same as Table 2. ``Transition 2'' refers to the $1s^{2}2s$
$^{2}S_{1/2}$$\longrightarrow$ $1s^{2}2p$ $^{2}P_{3/2}^{\circ}$ transition.}}}
\end{table}

%\subsection{Estimating the $d$-term}

We are now ready to show that it is possible to extract $\delta\left\langle r^{2}\right\rangle $ and
$\delta\left\langle r^{4}\right\rangle $ moments if accurate experimental field shifts are available.
This is due to the fact that the electronic factors $c_{k}$ and the $d_k$ terms can be accurately estimated
also when the rms radii is not known for the reference and/or target isotope. In these cases, we make instead
a ``qualified guess'' for the rms radii. The parametrization, given in Eq.~\ref{eq:grasp_formula}, for
the rms radius of an isotope $A$ constitutes an example of such a ``qualified guess'' and will be used below.
%The spherical
%Fermi model (see Sec. IV) can subsequently be used to predict the
%higher order radial moments. Thus, the differences $\delta\left\langle r^{2N}\right\rangle $
%for $N=1,..,4$ can be estimated. Assuming such nuclear parameters,
%variational calculations are performed with \pr{GRASP2K} in order
%to get the $\delta\nu_{VA}$ field shift values. Using the same radial
%moments differences the $\delta\nu_{RFS}$ field shifts are also calculated.
%The difference between $\delta\nu_{VA}$ and $\delta\nu_{RFS}$ gives
%an estimate of the $d$-term. This procedure relies on the shifts
%due to QED effects being relatively insensitive to small variations
%of the nuclear radius.

\subsubsection{Rms radii data available for the reference isotopes}

When radial moment differences are deduced from isotope shift measurements,
the nuclear parameters are usually known for the reference isotope but not for the target isotope.
We will now demonstrate the procedure for how experimental $\delta\left\langle r^{2}\right\rangle $ and
$\delta\left\langle r^{4}\right\rangle $ moments for the $^{238,234}$U isotope can be extracted in such cases by considering
the two resonance transitions in Li-like uranium. In what follows $^{238}$U is the reference isotope,
$r_A^{tab}$ denote a tabulated rms radius for isotope A taken from~\cite{Angeli}, $r_A^{para}$ denote
a parametrized rms radius for isotope A using Eq.~\ref{eq:grasp_formula} and spherical Fermi distributions
with $t=2.30$~fm is used everywhere. Further on it is assumed that accurate $\delta\nu_{k,FS}^{exp}$ values are available.
\begin{enumerate}
\item Two separate variational calculations are performed using $r_{238}^{tab}=5.8571$~fm and $r_{234}^{para}=5.7216$~fm, respectively.  
\item $\delta\nu_{k,VA}^{exact}$ is constructed using the level energies from the $r_{238}^{tab}$
  and $r_{234}^{para}$ calculations in step 1.
\item $\delta\nu_{k,RFS}$ is computed by using the electronic factors from the $r_{238}^{tab}$ calculation
  and the difference in radial moments as predicted by two spherical Fermi distributions with
  $r_{238}^{tab}$ and $r_{234}^{para}$, respectively.
\item $d_k=\delta\nu_{k,VA}^{exact}-\delta\nu_{k,RFS}$ is computed.
\item $c_k$ factors are computed using the electronic factors in step 3 (see Appendix A).
%\item $\delta\nu_{k,VA}^{exp}$ is constructed using the level energies from the $r_{238}^{tab}$
%  and $r_{234}^{tab}$ calculations in step 1.
\item $\delta\left\langle y_{1}\right\rangle$ and $\delta\left\langle y_{2}\right\rangle$ are extracted by solving
  Eq.~\ref{eq:theequation} and finally
\item $\delta\left\langle r^2\right\rangle$ and $\delta\left\langle r^4\right\rangle$ are computed (see Appendix A).  
\end{enumerate}

To quantatively validate the method we replace $\delta\nu_{k,FS}^{exp}$ with ``pseudo-experimental'' field shifts
constructed from two separate variational calculations using $r_{238}^{tab}$ and $r_{234}^{tab}=5.8291$~fm, respectively. In addition
we repeat the procedure for the $^{238,236}$U isotope pair using $r_{236}^{para}=5.7363$~fm and $r_{236}^{tab}=5.8431$~fm.
In Tab.~IV, the extracted $\delta\left\langle r^{2}\right\rangle $
and $\delta\left\langle r^{4}\right\rangle $ moments are compared
to the experimental $\delta\left\langle r^{2}\right\rangle _{exp}$ and $\delta\left\langle r^{4}\right\rangle _{exp}$ moments.
As seen, the extracted $\delta\left\langle r^{2}\right\rangle $
moments are almost identical to the ``experimental'' values. In addition
the $\delta\left\langle r^{4}\right\rangle $ moments are extracted
with an accuracy of $5.3$\% and $4.2$\% for the $^{234,238}$U and $^{236,238}$U pairs, respectively.
The errors, which are of systematical nature and remarkably small, arise from estimating the $d$-term using rms radii for the target isotopes
which differ by approximately 0.11~fm from the tabulated values used to conctruct the ``pseudo experimental''
field shifts. However, after the
extraction one obtains a better estimate for the rms radii of the target isotopes that
allows the method to be iteratively improved.

\begin{table}
\begin{tabular}{>{\raggedright}p{2cm}>{\raggedright}p{2cm}l}
\hline 
\noalign{\vskip0.1cm}
 & $\mathbf{238,234}$  & $\mathbf{238,236}$\tabularnewline
\hline 
\hline 
\noalign{\vskip0.1cm}
$\mathbf{\delta\left\langle \mathrm{r^{2}}\right\rangle }$  & -$0.3282$  & -$0.1642$\tabularnewline
$\mathbf{\delta\left\langle \mathrm{r^{2}}\right\rangle _{exp}}$  & -$0.3272$  & -$0.1638$\tabularnewline
$error$ & $\mathbf{0.0010}$  & $\mathbf{0.0004}$\tabularnewline
\noalign{\vskip0.1cm}
$\mathbf{\mathbf{\delta\left\langle \mathrm{r^{4}}\right\rangle }}$  & -$28.9026$  & -$14.3453$\tabularnewline
$\mathbf{\mathbf{\mathbf{\delta\left\langle \mathrm{r^{4}}\right\rangle }}_{exp}}$  & -$27.4419$  & -$13.7693$\tabularnewline
$error$ & $\mathbf{1.4607}$  & $\mathbf{0.5760}$\tabularnewline
\hline 
\end{tabular}\caption{{\small{Errors, in fm$^2$ and fm$^4$, when extracting the $\delta\left\langle r^{2}\right\rangle $
and $\delta\left\langle r^{4}\right\rangle $ moments, for the $^{234,238}$U
and $^{236,238}$U pairs. It is assumed that the rms radii are unknown for the target isotopes. See text for details.}}}
\end{table}

\subsubsection{Rms radii unknown for both target and reference isotopes}

Assuming that the rms radius value of the reference isotope is also
unknown, we again try to extract the $\delta\left\langle r^{2}\right\rangle $
and $\delta\left\langle r^{4}\right\rangle $ moments. A ``qualified
guess'' for the rms radius of $^{238}$U is then needed and we replace $r_{238}^{tab}$ with $r_{238}^{para}=5.7508$~fm
in the procedure described above.

The results from the extraction of the $\delta\left\langle r^{2}\right\rangle $
and $\delta\left\langle r^{4}\right\rangle $ moments are presented
in Tab.~V. As seen the $\delta\left\langle r^{2}\right\rangle $
moment is extracted almost as accurate as before (see Tab.~IV). Further
on, the results from extracting the $\delta\left\langle r^{4}\right\rangle $
moments display a discrepancy of $\sim10.3$\% and $6.5$\% from the
exact values, for the $^{234,238}$U and $^{236,238}$U pairs respectively. 

The nuclear parameters relevant to the reference isotope have been
modified here. Thus, the $F_{k}$ factors have also been re-evaluated
since they are always deduced for the reference isotope. As a result,
besides the new radial moments differences, the $\delta\nu_{k,RFS}$
field shifts are computed based on updated sets of $F_{k,N}$ factors.
This explains the larger discrepancy that is observed when extracting
the $\delta\left\langle r^{2}\right\rangle $ and $\delta\left\langle r^{4}\right\rangle $
moments in the latter case (see Tab.~V). However, the results are remarkably good given
that the ``qualified guess'' for the reference isotope is approximately 0.11~fm smaller than
the tabulated value used to construct the ``pseudo experimental'' field shifts. 

\begin{table}
\begin{tabular}{>{\raggedright}p{2cm}>{\raggedright}p{2cm}l}
\hline 
\noalign{\vskip0.1cm}
 & $\mathbf{\mathbf{238,234}}$  & $\mathbf{\mathbf{238,236}}$\tabularnewline
\hline 
\hline 
\noalign{\vskip0.1cm}
$\mathbf{\delta\left\langle \mathrm{r^{2}}\right\rangle }$  & -$0.3287$  & -$0.1640$\tabularnewline
$\mathbf{\delta\left\langle \mathrm{r^{2}}\right\rangle _{exp}}$  & -$0.3272$  & -$0.1638$\tabularnewline
$error$ & $\mathbf{0.0015}$  & $\mathbf{0.0002}$\tabularnewline
\noalign{\vskip0.1cm}
$\mathbf{\mathbf{\delta\left\langle \mathrm{r^{4}}\right\rangle }}$  & -$30.2665$  & -$14.6612$\tabularnewline
$\mathbf{\mathbf{\mathbf{\delta\left\langle \mathrm{r^{4}}\right\rangle }}_{exp}}$  & -$27.4419$  & -$13.7693$\tabularnewline
$error$ & $\mathbf{2.8246}$  & $\mathbf{0.8919}$\tabularnewline
\hline 
\end{tabular}\caption{{\small{{}Same as Table IV. Here it is assumed that rms radii are unknown for both
      the reference and the target isotopes. See text for details.}}}
\end{table}

\subsection{Statistical errors when extracting the $\delta\left\langle r^{2}\right\rangle $
and $\delta\left\langle r^{4}\right\rangle $ moments}

Above, the $\delta\left\langle r^{2}\right\rangle $ and $\delta\left\langle r^{4}\right\rangle $
moments were extracted by solving the matrix equation: 
\begin{equation}\label{eq:matrixeq}
\left[\begin{array}{c}
\delta\nu_{1,RFS}\\
\delta\nu_{2,RFS}
\end{array}\right] = C\left[\begin{array}{c}
\delta\left\langle y_{1}\right\rangle \\
\delta\left\langle y_{2}\right\rangle 
\end{array}\right].
\end{equation}
In order to solve for $y_{1}$ and $y_{2}$, the matrix $C$ must
be invertible. If the matrix determinant is zero, then the matrix
is singular and cannot be inverted. It is not rare that the determinant
of such matrix can be close to zero, but still non-zero. In this case,
the matrix is close to singular and as a result the values of $\delta\left\langle y_{1}\right\rangle $
and $\delta\left\langle y_{2}\right\rangle $ will be hugely affected,
even by a small change in the field shifts $\delta\nu_{1,RFS}$ and
$\delta\nu_{2,RFS}$. Namely, the extracted $\delta\left\langle y_{1}\right\rangle $
and $\delta\left\langle y_{2}\right\rangle $ values, and as a consequence
the $\delta\left\langle r^{2}\right\rangle $ and $\delta\left\langle r^{4}\right\rangle $
moments, will to a great degree be affected by the uncertainties in
the observed isotope shifts, making the extraction of the radial nuclear
moments with high accuracy a difficult task. A $C$ matrix determinant
equal to zero is obtained if the two equations are linearly dependent.
In such case it is not possible to extract two unknowns. Therefore,
the transitions considered should be as independent as possible in
terms of electronic factors.

The observed isotope shifts $\delta\nu_{k,IS}^{exp}$, and subsequently
the observed field shifts $\delta\nu_{k,FS}^{exp}$, are associated with uncertainties of a certain magnitude. These
uncertainties lead to statistical errors in the extracted nuclear moments.
In the next subsections the propagation of these errors is
discussed and how they can be minimized by selecting atomic transitions.

\subsubsection{Statistical errors in relation to the atomic number}

In Subsec.~B, our new method was tested by using $\delta\nu_{RFS}$
line field shifts as ``pseudo experimental'' data. In order to extend this approach
to consider uncertainties we assume uncorrelated errors with an uncertainty
$\pm\epsilon$, where $\epsilon=\delta\nu_{k,RFS}\cdot10^{-m}$, in
the $\delta\nu_{k,RFS}$ values that are used for solving the matrix
equation (see Eq.~\ref{eq:matrixeq}). By varying $m$, the magnitude of the field
shift uncertainty changes. We can then investigate the effect these
uncertainties have on the extracted $\delta\left\langle r^{2}\right\rangle $
and $\delta\left\langle r^{4}\right\rangle $ values.

The extraction of the $\delta\left\langle r^{2}\right\rangle $ and
$\delta\left\langle r^{4}\right\rangle $ moments was in Subsec.~B
performed for several uranium, lead and neodymium isotope pairs (see
Fig.~10). By making a reasonable choice of $m=3$ for the error $\epsilon$
in the $\delta\nu_{k,RFS}$ values relevant to these isotope pairs,
it is possible to estimate the magnitude of the statistical errors in the
extracted $\delta\left\langle r^{2}\right\rangle $ and $\delta\left\langle r^{4}\right\rangle $
moments. The relative errors of the extracted values for one isotope
pair of each of the above elements, are indicatively presented in
Tab.~VI. The error in $\delta\left\langle r^{2}\right\rangle ^{142,150}$
is approximately $72\%$ of the magnitude of the resulting value.
Besides, the $\delta\left\langle r^{4}\right\rangle ^{142,150}$ is
extracted with significantly greater error. However, the relative
error in both $\delta\left\langle r^{2}\right\rangle $ and $\delta\left\langle r^{4}\right\rangle $
demonstrates a considerable decrease as the atomic number of the isotopes
becomes larger.

\begin{table}
\begin{tabular}{>{\centering}p{1.5cm}>{\raggedleft}p{1.8cm}>{\raggedleft}p{1.8cm}>{\raggedleft}p{1.8cm}}
\hline 
\noalign{\vskip0.1cm}
 & {\small{$\mathbf{^{142,150}Nd}$}}  & {\small{$\mathbf{^{208,192}Pb}$}}  & {\small{$\mathbf{^{238,236}U}$}}\tabularnewline
\noalign{\vskip0.1cm}
{\large{$Z$}}  & $60$  & $82$  & $92$\tabularnewline[0.15cm]
\hline 
\hline 
\noalign{\vskip0.1cm}
{\large{$\frac{\Delta(\delta\left\langle r^{2}\right\rangle )}{\left|\delta\left\langle r^{2}\right\rangle \right|}$}}  & $0.72$  & $0.39$  & $0.28$\tabularnewline[0.15cm]
\noalign{\vskip0.1cm}
{\large{$\frac{\Delta(\delta\left\langle r^{4}\right\rangle )}{\left|\delta\left\langle r^{4}\right\rangle \right|}$}}  & $13.84$  & $5.54$  & $3.65$\tabularnewline[0.15cm]
\hline 
\end{tabular}\caption{{\small{{}The relative error in the extraction of the $\delta\left\langle r^{2}\right\rangle $
and $\delta\left\langle r^{4}\right\rangle $ moments for the $^{142,150}$Nd$^{57+}$,
$^{208,192}$Pb$^{79+}$ and $^{238,236}$U$^{89+}$ pairs. The relative
errors are presented as a function of the atomic number of these three
elements. The inaccuracy assumed in the $\delta\nu_{k,RFS}$ field
shift data is $\pm\epsilon=\delta\nu_{k,RFS}\cdot10^{-3}$.}}}
\end{table}

\medskip{}
So far, the extraction of the $\delta\left\langle r^{2}\right\rangle $
and $\delta\left\langle r^{4}\right\rangle $ moments was performed
by making use of $\delta\nu_{k,RFS}$ field shifts and $F_{k,N}$ line
field shift factors that are attributed to the first two resonance
transitions, i.e. $1s^{2}2s$ $^{2}S_{1/2}$$\longrightarrow$ $1s^{2}2p$
$^{2}P_{1/2}^{\circ}$ and $1s^{2}2s$ $^{2}S_{1/2}$$\longrightarrow$
$1s^{2}2p$ $^{2}P_{3/2}^{\circ}$. For these two transitions in lithium-like
systems, the $F_{k,N}$ factors, as well as the line mass shift parameters
$\Delta K_{k,MS}$, can be determined with high accuracy.
Therefore, when we in practice attempt to extract the $\delta\left\langle r^{2}\right\rangle $
and $\delta\left\langle r^{4}\right\rangle $ moments using actual
experimental data, the uncertainties in the $\delta\nu_{k,FS}^{exp}$
values will normally be dominated by the uncertainties in the
$\delta\nu_{k,IS}^{exp}$ measurements. 

For the $^{142,150}$Nd$^{57+}$ pair and the previously mentioned
transitions such measurements are available \cite{Exp_IS_ND57+}.
Taking into account the uncertainties in the measured 
isotope shifts $\delta\nu_{k,IS}^{exp}$, the corresponding
uncertainties in $\delta\nu_{k,FS}^{exp}$ appear in the fourth
and third digit for each of the above transitions, respectively. In
this case the choice of an error $\pm\epsilon=\delta\nu_{k,RFS}\cdot10^{-3}$
in the calculated field shift values seems to be quite realistic.
However, according to Tab.~VI the errors in the $\delta\left\langle r^{2}\right\rangle $
and $\delta\left\langle r^{4}\right\rangle$ values resulting from experimental uncertainties
of this magnitude for the neodymium pair are evidently extremely
large. 

We can therefore draw the conclusion that the extraction of the $\delta\left\langle r^{2}\right\rangle ^{142,150}$
and $\delta\left\langle r^{4}\right\rangle ^{142,150}$ moments with
satisfactory accuracy is not likely to be a possibility at the moment.
Varying $m$ we deduce that in order for the $\delta\left\langle r^{2}\right\rangle ^{142,150}$
and $\delta\left\langle r^{4}\right\rangle ^{142,150}$ to be determined
with uncertainties of the order of $\lesssim1$\% and $\lesssim14$\%,
respectively, we should assume $m\geq5$. In addition, considering
Tab.~VI, a more precise extraction of the $\delta\left\langle r^{2}\right\rangle $
and $\delta\left\langle r^{4}\right\rangle $ moments should be possible
for the lead and in particular for the uranium isotope pairs.

\subsubsection{Independent transitions}

Considering the two resonance transitions that were used above for
extracting $\delta\left\langle r^{2}\right\rangle $ and $\delta\left\langle r^{4}\right\rangle $
moments, we note that the same final state
takes part in both. Therefore, these two transitions are not entirely
independent and the corresponding $F_{k,N}$ factors do not constitute
the best possible set so that we avoid matrix $C$ being close to
singular. As a consequence, the uncertainties in the $\delta\left\langle r^{2}\right\rangle $
and $\delta\left\langle r^{4}\right\rangle $ values are relatively large.
%Even so, in case we are willing to extract only the $\delta\left\langle r^{2}\right\rangle $
%moments and therefore solving the matrix Eq.~\ref{eq:matrixeq} for one unknown,
%the statistical errors vanish. In this case, the extracted $\delta\left\langle r^{2}\right\rangle $
%moments will only suffer from model errors which are now greater due
%to the fact that only the first expansion term of Eq.~\ref{eq:yeq} is considered. 
In order to be able to accurately extract both $\delta\left\langle r^{2}\right\rangle $
and $\delta\left\langle r^{4}\right\rangle $ moments, the precision
of the experimental methods must therefore be improved substantially. Alternatively,
a larger number of transitions must be available. Using the \pr{GRASP2K}
package we can easily compute line field shift parameters for more transitions and
hence an extended set of $\delta\nu_{k,RFS}$ values can be generated. The
matrix equation will then be formed using $k>2$ equations, which
need to be solved for the same unknowns, $y_{1}$ and $y_{2}$. Having
more equations than number of unknowns leads to a reduction of the
statistical errors. 

Choosing, for instance, to extract the $\delta\left\langle r^{2}\right\rangle $
and $\delta\left\langle r^{4}\right\rangle $ moments for the $^{238,236}$U
pair, we solve a matrix equation that consists of 16 equations corresponding
to 16 different transitions. These transitions involve the following
even: $1s^{2}2s$ $^{2}S_{1/2}$, $\,1s^{2}3s$ $^{2}S_{1/2}$, $\,1s^{2}3d$
$^{2}D_{3/2,5/2}$ and odd: $1s^{2}2p$ $^{2}P_{1/2,3/2}^{\circ}$, $\,1s^{2}3p$
$^{2}P_{1/2,3/2}^{\circ}$ states in Li-like uranium. By making the same choice of $m=3$ for the error $\epsilon=\delta\nu_{k,RFS}\cdot10^{-m}$
in the $\delta\nu_{k,RFS}$ values, we extract the $\delta\left\langle r^{2}\right\rangle $
and $\delta\left\langle r^{4}\right\rangle $ moments. The extracted
$\delta\left\langle r^{2}\right\rangle $ moment has exactly the same
value as before, whereas the $\delta\left\langle r^{4}\right\rangle $
value is also about the same, suffering from approximately the same
systematical errors. However, the statistical errors in the extraction of
both $\delta\left\langle r^{2}\right\rangle $ and $\delta\left\langle r^{4}\right\rangle $
have now been decreased significantly (see Tab.~VII).

\begin{table}
\begin{tabular}{>{\centering}p{1.8cm}r>{\raggedleft}p{1.3cm}>{\raggedleft}p{1.5cm}}
\hline 
\noalign{\vskip0.1cm}
 {\small{$\mathbf{^{238,236}U}$}}  & 2 res. & all 16  & 2 ind.\tabularnewline[0.1cm]
\hline 
\hline 
\noalign{\vskip0.1cm}
{\large{$\frac{\Delta(\delta\left\langle r^{2}\right\rangle )}{\left|\delta\left\langle r^{2}\right\rangle \right|}$}}  & $0.28$  & $0.03$  & $0.02$\tabularnewline[0.1cm]
\noalign{\vskip0.1cm}
{\large{$\frac{\Delta(\delta\left\langle r^{4}\right\rangle )}{\left|\delta\left\langle r^{4}\right\rangle \right|}$}}  & $3.65$  & $0.38$  & $0.30$\tabularnewline[0.14cm]
\hline 
\end{tabular}\caption{{\small{{}The relative error in the extraction of the $\delta\left\langle r^{2}\right\rangle $
and $\delta\left\langle r^{4}\right\rangle $ moments for the $^{238,236}$U$^{89+}$
pair, initially calculated when the first two resonance transitions
were studied, when all 16 theoretically available transitions are
used and when we finally choose one set of as independent as possible
transitions. The uncertianties assumed in the $\delta\nu_{k,RFS}$ field
shift data is, as in Tab.~VI, $\pm\epsilon=\delta\nu_{k,RFS}\cdot10^{-3}$.}}}
\end{table}

In practice, such large number of measured transitions is not likely
to be available. Trying all different combinations we realize that the
error in the extraction of the $\delta\left\langle r^{2}\right\rangle $
and $\delta\left\langle r^{4}\right\rangle $ moments, by using a
set of only two transitions, varies with the choice of the transitions.
For the $^{238,236}$U pair and $\epsilon=\delta\nu_{k,RFS}\cdot10^{-3}$
we get $0.0014\leq\frac{\Delta(\delta\left\langle r^{2}\right\rangle )}{\left|\delta\left\langle r^{2}\right\rangle \right|}\leq80$
and $0.0012\leq\frac{\Delta(\delta\left\langle r^{4}\right\rangle )}{\left|\delta\left\langle r^{4}\right\rangle \right|}\leq1100$,
for the relative errors in the extraction of the $\delta\left\langle r^{2}\right\rangle $
and $\delta\left\langle r^{4}\right\rangle $ moments, respectively.

We therefore deduce that in order to limit the magnitude of the statistical
errors, it is more important to make a choice of as independent
as possible transitions that form the set of equations solved, rather
than increasing the number of transitions. Based on this conclusion,
instead of extracting the $\delta\left\langle r^{2}\right\rangle $
and $\delta\left\langle r^{4}\right\rangle $ moments using the first
two resonance transitions, a set of two more independent transitions
is chosen. Thus, we attempt to extract the $\delta\left\langle r^{2}\right\rangle $
and $\delta\left\langle r^{4}\right\rangle $ moments for the $^{238,236}$U
isotope pair, using the resonance transition $1s^{2}2s$ $^{2}S_{1/2}$$\longrightarrow$
$1s^{2}2p$ $^{2}P_{1/2}^{\circ}\,$ combined with the $\,1s^{2}3p$ $^{2}P_{1/2}^{\circ}$
$\longrightarrow$ $1s^{2}3d$ $^{2}D_{3/2}\,$ transition. The resulting
relative errors for this combination of transitions are also displayed
in Tab.~VII. As seen, the relative errors in the extraction of both
$\delta\left\langle r^{2}\right\rangle $ and $\delta\left\langle r^{4}\right\rangle $
moments are decreased when a more optimal combination of two out of
the total 16 available transitions is chosen.

\begin{table}
\begin{tabular}{>{\raggedright}p{2cm}>{\raggedright}p{2cm}l}
\hline 
\noalign{\vskip0.1cm}
\multicolumn{3}{l}{$\mathbf{\mathbf{\qquad\qquad\qquad\,\,238,236}}$}\tabularnewline
\hline 
\hline 
\noalign{\vskip0.1cm}
$\mathbf{\delta\left\langle \mathrm{r^{2}}\right\rangle }$  &  -$0.1646$ & $\pm0.0036$\tabularnewline
$\mathbf{\delta\left\langle \mathrm{r^{2}}\right\rangle }_{\mathbf{exact}}$  & -$0.1638$  & -\tabularnewline
$error$  & $\mathbf{0.0008}$  & -\tabularnewline
\noalign{\vskip0.1cm}
$\mathbf{\delta\left\langle \mathrm{r^{4}}\right\rangle }$  & -$14.7283$ & $\pm3.5279$\tabularnewline
$\mathbf{\delta\left\langle \mathrm{r^{4}}\right\rangle }_{\mathbf{exact}}$  & -$13.7693$  & -\tabularnewline
$error$  & $\mathbf{0.9590}$  & -\tabularnewline
\hline 
\end{tabular}\caption{{\small{{}Same as Tab.~IV. Here, the line field shift factors $F_{k,N}$
correspond to the $1s^{2}2s$ $^{2}S_{1/2}$$\longrightarrow$ $1s^{2}2p$
$^{2}P_{1/2}^{\circ}$ and $1s^{2}3p$ $^{2}P_{1/2}$ $\longrightarrow$
$1s^{2}3d$ $^{2}D_{3/2}$ transitions. Statistical errors are given in the rightmost column assuming
uncertainties in the ``pseudo experimental'' field shifts according to $\epsilon=\delta\nu_{k,RFS}\cdot10^{-3}$.}}}
\end{table}

\begin{figure}
\includegraphics[width=0.85\columnwidth]{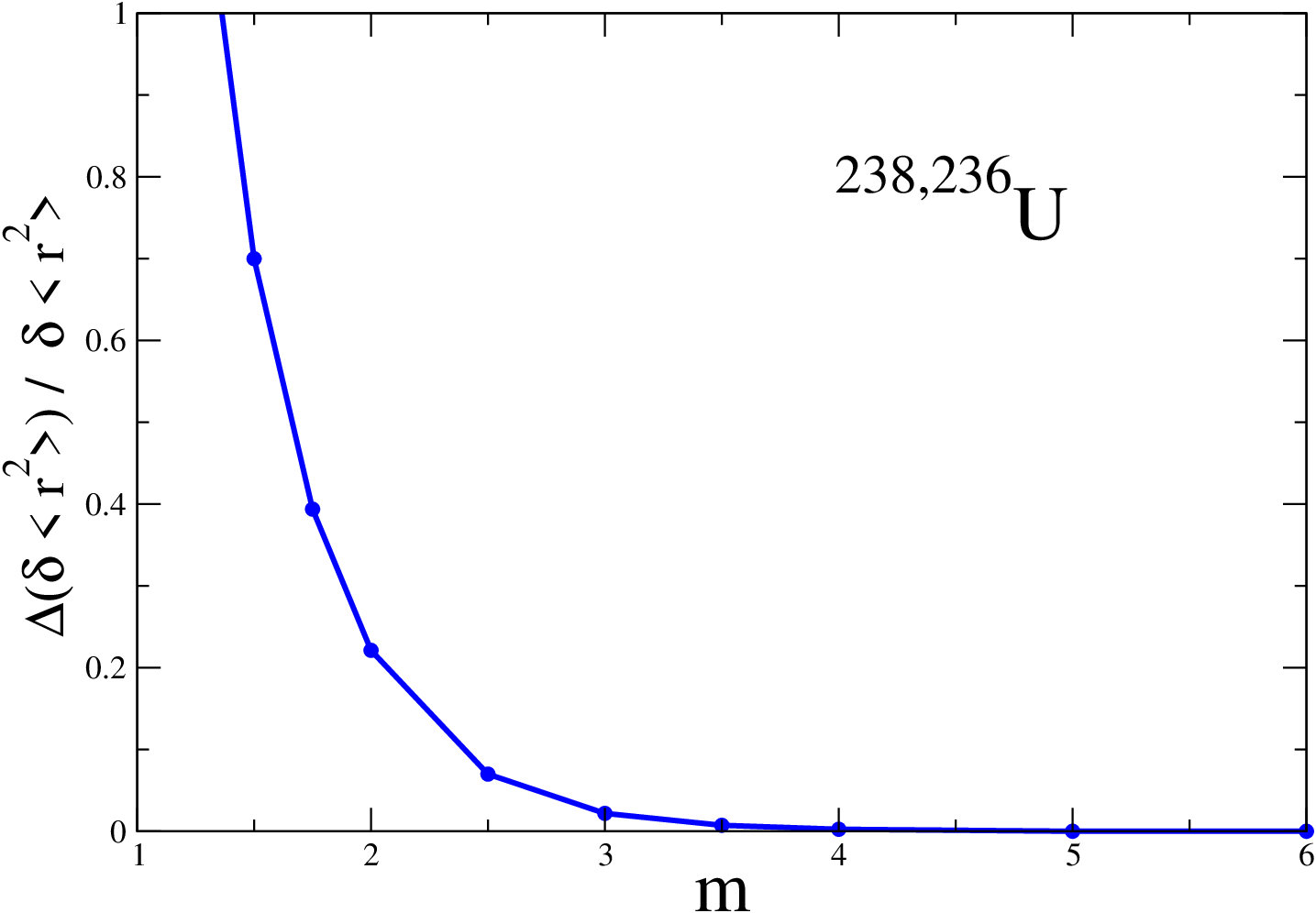}

\caption{{\small{{}The relative error in the extraction of the $\delta\left\langle r^{2}\right\rangle $
moment as a function of $m$ number in the assumed uncertainty
$\epsilon=\delta\nu_{k,RFS}\cdot10^{-m}$ of the field shift. For the extraction, the pair of $1s^{2}2s$ $^{2}S_{1/2}$$\longrightarrow$
$1s^{2}2p$ $^{2}P_{1/2}^{\circ}$ and $1s^{2}3p$ $^{2}P_{1/2}$ $\longrightarrow$
$1s^{2}3d$ $^{2}D_{3/2}$ transitions in Li-like $^{238,236}$U has been used.}}}
\end{figure}

\subsubsection{Errors in the extraction of $\delta\left\langle r^{4}\right\rangle ^{238,236}$}

Having ascertained that the ``right'' combination of transitions
provides us with reasonably small statistical errors, we can extract
the $\delta\left\langle r^{2}\right\rangle $ and $\delta\left\langle r^{4}\right\rangle $
moments for the $^{238,236}$U isotope pair using ``pseudo-experimental''
field shifts, as described in Subsec. C1, for this ``optimal'' pair of transitions.
%The $d_{k}$-terms are estimated by making a ``qualified guess''
%for the rms radius of the target isotope $^{236}$U according to Subsec.
%D.
The statistical uncertainties are estimated as $\epsilon=\delta\nu_{k,RFS}\cdot10^{-m}$ with $m=3$, which has
been used so far for determining the assumed uncertainty in the $\delta\nu_{k,RFS}$ values. 

The extracted radial moments together with the resulting errors are displayed
in Tab.~VIII. Comparing the respective results of Tab.~IV with the
results in Tab.~VIII, we deduce that although in the latter case
the systematical errors are larger the statistical errors of the extracted
$\delta\left\langle r^{2}\right\rangle $ and $\delta\left\langle r^{4}\right\rangle $
values are significantly smaller. We see that now the relative statistical
errors are $\frac{\Delta(\delta\left\langle r^{2}\right\rangle )}{\left|\delta\left\langle r^{2}\right\rangle \right|}=0.022$
and $\frac{\Delta(\delta\left\langle r^{4}\right\rangle )}{\left|\delta\left\langle r^{4}\right\rangle \right|}=0.24$,
respectively. 

In Fig.~11 and 12, the relative errors in the extraction of the $\delta\left\langle r^{2}\right\rangle $
and $\delta\left\langle r^{4}\right\rangle $ moments are illustrated
as a function of the $m$ value. As seen the results are rather sensitive
to the $m$ value and the  relative error increases dramatically as
the precision of the field shift values decreases. This is even more pronounced
for the errors in the extracted $\delta\left\langle r^{4}\right\rangle $
moments. Nevertheless, for $m=3$ both $\delta\left\langle r^{2}\right\rangle $
and $\delta\left\langle r^{4}\right\rangle $ moments are extracted
with satisfactory accuracy. Thus we deduce, that provided the current
experimental precision in the isotope shift measurements, an accurate
enough extraction of the $\delta\left\langle r^{2}\right\rangle $
and $\delta\left\langle r^{4}\right\rangle $ moments could be possibile
as long as the measured transitions are sufficiently independent in terms of electronic factors. 

\begin{figure}
\includegraphics[width=0.85\columnwidth]{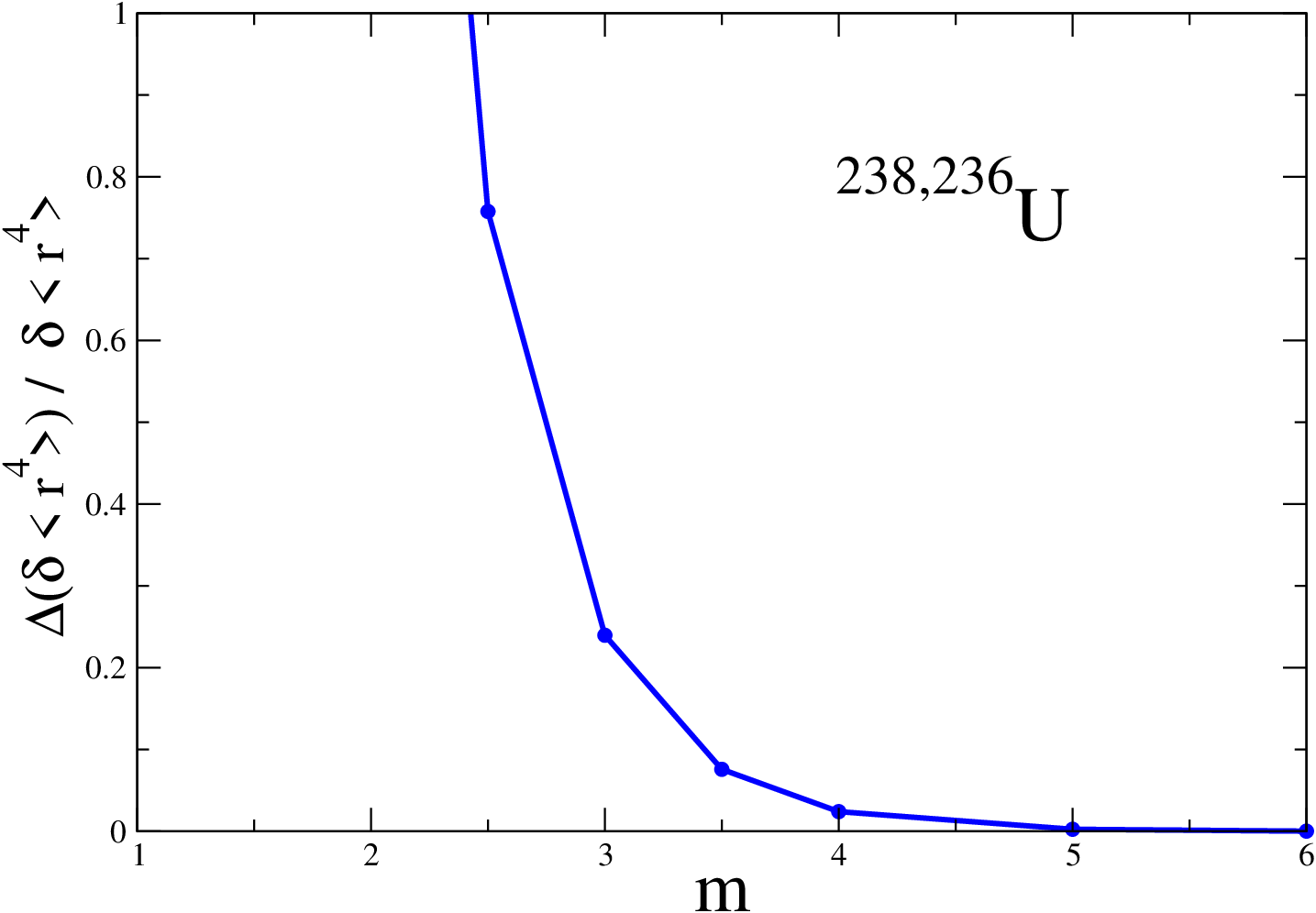}\caption{{\small{{}Same as Fig.~11, but for the relative uncertainty in the
extraction of the $\delta\left\langle r^{4}\right\rangle $ moment
using the same pair of transitions.}}}
\end{figure}

\section{summary and conclusions}

Combining nuclear DFT-type models with MCHF calculations for atomic
states it is possible to achieve a higher precision in the predictions
of atomic line field shifts. Changes in the nuclear charge distribution
caused by shell structure, deformations and variations in the diffusseness
of the nuclei are then automatically taken into account. In this work
it is shown that capturing all these effects leads to an improved
description of experiments. 

With the continous advancement in experimental methods one may ask
whether the improved precision and access to several atomic transitions
makes it possible to obtain more data on the nuclear isotopes than
just the $\delta\left\langle r^{2}\right\rangle $ values commonly
extracted so far. By constructing a set of theoretical field shifts
we explore the possibility of extracting information about the nucleus
by inverting the first order perturbation theory equations for the
field shifts. In this way we demonstrate that the electron states
are sensitive not only to the $\delta\left\langle r^{2}\right\rangle $
values but also to changes in $\left\langle r^{4}\right\rangle $
values. This opens the possibility for systematic tabulation of these
higher order nuclear moments. Considering both statistical and systematical
errors in the extraction procedure we conclude that an increase in
experimental precision by 1-2 orders of magnitude or access to data
for more independent atomic transitions would be essential. As a promising
candidate for future experiments we suggest Li-like Uranium where
an increase in precision with one order of magnitude along with access
to at least two independent transitions would allow accurate $\delta\left\langle r^{4}\right\rangle $
values to be extracted.

\section*{acknowledgments}

This work was supported by the Swedish Research Council (VR) (grant number:
2015-04842). The authors acknowledge Per Jönsson, Malmö University, for discussions
and C. Nazé, J.G. Li and M. Godefroid for providing the barium MCDHF wave-functions.

\bigskip{}

\clearpage{}\newpage{}

\bibliographystyle{unsrt}
\bibliography{references}

\begin{thebibliography}{10}

\bibitem{Angeli}
I.~Angeli and K.~Marinova.
\newblock {\em At. Data Nucl. Data Tables}, 99:69, (2013).

\bibitem{Hydr_Deut}
C.G.~Parthey at~al.
\newblock {\em Phys. Rev. Lett.}, 104:233001, (2010).

\bibitem{Agbemava2014}
S.~E. Agbemava, A.~V. Afanasjev, D.~Ray, and P.~Ring.
\newblock {\em Phys. Rev. C}, 89:054320, 2014.

\bibitem{Andrae}
D.~Andrae.
\newblock {\em Phys. Rep.}, 336:413, (2000).

\bibitem{Dielectr_exp}
Y.~Kozhedub, O.~Andreev, V.~Shabaev, I.~Tupitsyn, C.~Brandau, C.~Kozhuharov,
  G.~Plunien, and T.~Stöhlker.
\newblock {\em Phys. Rev. A}, 77:032501, (2008).

\bibitem{Zubova}
N.A.~Zubova et~al.
\newblock {\em Phys. Rev. A}, 90:062512, (2014).

\bibitem{ISOLDE}
C.~Brandau et~al.
\newblock {\em J. Phys. Conf. Ser.}, 635:022084, (2015).

\bibitem{GSI}
M.J.G Borge.
\newblock {\em J. Phys. Conf. Ser.}, 580:012049, (2015).

\bibitem{Mass_shift_ref3}
C.~W.~P Palmer.
\newblock {\em J. Phys. B: At. Mol. Phys.}, 20:5987, (1987).

\bibitem{nuclear_recoil_corr}
V.M. Shabaev.
\newblock {\em Theor. Math. Phys.}, 63:588, (1985).

\bibitem{Nuclear_recoil_corr_ref4}
V.~Shabaev and A.~Artemyev.
\newblock {\em J. Phys. B: At. Mol. Phys.}, 27:1307, (1994).

\bibitem{Isotope_shifts}
E.C. Seltzer.
\newblock {\em Phys. Rev.}, 188:1916, (1969).

\bibitem{e_density}
G.~Torbohm, B.~Fricke, and A.~Rosén.
\newblock {\em Phys. Rev. A}, 31, (1985).

\bibitem{e_density_2}
S.~Blundell, P.~Baird, C.~Palmer, D.~Stacey, and G.~Woodgate.
\newblock {\em J. Phys. B: At. Mol. Phys.}, 20:3663, (1987).

\bibitem{tobepublished}
J.~Ekman et~al.
\newblock {\em Submitted to Comp. Phys. Comm.}

\bibitem{Grasp2K_2013}
P.~Jönsson, G.~Gaigalas, J.~Bieroń, C.~Froese Fischer, and I.~Grant.
\newblock {\em Comput. Phys. Commun.}, 184:2197, (2013).

\bibitem{Relat_atoms}
I.~P. Grant.
\newblock {\em Springer Series on Atomic, Optical and Plasma Physics}, 40,
  (2007).

\bibitem{Breit_QED}
B.J. McKenzie, I.P. Grant, and P.H. Norrington.
\newblock {\em Comp. Phys. Comm.}, 21:233, (1980).

\bibitem{ris3u}
C.~Nazé, E.~Gaidamauskas, G.~Gaigalas, M.~Godefroid, and P.~Jönsson.
\newblock {\em Comput. Phys. Commun.}, 184:2187, (2013).

\bibitem{Ring_Schuck}
P.~Ring and P.~Schuck.
\newblock The nuclear many-body problem, 1st ed. springer-verlag, new york,
  (1980).

\bibitem{SLY4}
E.~Chabanat, P.~Bonche, P.~Haensel, J.~Meyer, and R.~Schaeffer.
\newblock {\em Nucl. Phys. A}, 635:231, (1998).

\bibitem{UNEDF1}
B.~Alex Brown.
\newblock {\em Phys. Rev. C}, 58:220, (1998).

\bibitem{Carlsson_hosphe_v1.02}
B.G. Carlsson, J.~Dobaczewski, J.~Toivanen, and P.~Veselý.
\newblock {\em Comput. Phys. Commun.}, 181:1641, (2010).

\bibitem{HFBTHO}
M.~Stoitsov, N.~Schunck, M.~Kortelainen, N.~Michel, H.~Nam, et~al.
\newblock {\em Comput. Phys. Commun.}, 184:1592, (2013).

\bibitem{Folding}
A.~Bouyssy, J.~Mathiot, N.~Van Giai, and S.~Marcos.
\newblock {\em Phys. Rev. C}, 36:380, (1987).

\bibitem{proton_radius0.88}
R.~Rosenfelder.
\newblock {\em Phys. Lett. B}, 479:381, 2000.

\bibitem{proton_radius}
R.~Pohl, R.~Gilman, G.~Miller, and K.~Pachucki.
\newblock {\em Annu. Rev. Nucl. Part. Sci.}, 63:175, (2013).

\bibitem{Exper_data}
H.~De Vries, C.W.~De Jager, and C.~De Vries.
\newblock {\em At. Data Nucl. Data Tables}, 36:495, (1987).

\bibitem{Ba_IS_measurements}
W.~van Wijngaarden and J.~Li.
\newblock {\em Can. J. Phys.}, 73:484, (1995).

\bibitem{Barium_Theoret_IS}
C.~Naz\'e, J.G. Li, and M.~Godefroid.
\newblock {\em Phys. Rev. A}, 91:032511, (2015).

\bibitem{Dielectr_exp_2}
N.~A. Zubova, Y.~S. Kozhedub, V.~M. Shabaev, I.~I. Tupitsyn, A.~V. Volotka,
  G.~Plunien, C.~Brandau, and T.~Stöhlker.
\newblock {\em Phys. Rev. A}, 90:062512, (2014).

\bibitem{li2012mass}
J.~Li, C.~Naz{\'e}, M.~Godefroid, S.~Fritzsche, G.~Gaigalas, P.~Indelicato, and
  P.~J{\"o}nsson.
\newblock {\em Physical Review A}, 86:022518, (2012).

\bibitem{Exp_IS_ND57+}
C.~Brandau et~al.
\newblock {\em Phys. Rev. Lett.}, 100:073201, (2008).

\bibitem{Elliott}
S.~R. Elliott, P.~Beiersdorfer, and M.~H. Chen.
\newblock {\em Phys. Rev. Lett.}, 76:1031--1034, (1996).

\bibitem{Gram-Schmidt_process}
Eric~W. Weisstein.
\newblock Gram-schmidt orthonormalization. from mathworld--a wolfram web
  resource.

\end{thebibliography}

\clearpage{}

\newpage{}

\appendix
%dummy comment inserted by tex2lyx to ensure that this paragraph is not empty

\section{RFS expansion in orthonormal basis}

\label{app-A}

The RFS is, for a certain transition, given by the expansion: 
\[
\sum_{N=1}^{4}F_{N}\delta\left\langle r^{2N}\right\rangle =F_{1}\delta\left\langle r^{2}\right\rangle +F_{2}\delta\left\langle r^{4}\right\rangle +F_{3}\delta\left\langle r^{6}\right\rangle +F_{4}\delta\left\langle r^{8}\right\rangle ,
\]
where the line field shift factors $F_{N}$ play the role of expansion
coefficients. The set of $r^{2N}$ that forms the basis $\left\{ r^{2},r^{4},r^{6},r^{8}\right\} $
is not orthonormal. It is reasonable to assume that a re-arrangement
using an orthonormal basis should lead to faster convergence. Here,
we orthonormalize the initial basis with respect to the scalar product:
\[
\left\langle u\mid v\right\rangle ={\displaystyle \int u*v*wr^{2}dr},
\]
where $w$ is the weight function that approximates the nucleus. Since
the functions $y_{N}$, forming the basis $\left\{ y_{1},y_{2},y_{3},y_{4}\right\} $,
are constructed to be orthogonal they will probe different aspects
of the nuclear charge distribution within the nuclear volume. Thus,
we expect that the new expansion: 
\[
\sum_{N=1}^{4}c_{N}\delta\left\langle y_{N}\right\rangle =c_{1}\delta\left\langle y_{1}\right\rangle +c_{2}\delta\left\langle y_{2}\right\rangle +c_{3}\delta\left\langle y_{3}\right\rangle +c_{4}\delta\left\langle y_{4}\right\rangle 
\]
will converge faster than ${\textstyle {\displaystyle {\textstyle \sum}_{N=1}^{4}}}F_{N}\delta\left\langle r^{2N}\right\rangle $
does. In the expression above, $c_{N}$ are the new expansion coefficients.
Assuming that the nucleus can be approximated as a hard sphere, one
can use $w=\rho_{0}\Theta(R-r)$ with $R=1.25A^{1/3}$. The value
of $\rho_{0}$ is determined by the normalization condition: $4\pi\int\rho_{0}r^{2}dr=1$.
Following the Gram-Schmidt process \cite{Gram-Schmidt_process}, we
obtain: 
\[
y_{1}=\frac{3.46556}{\bar{A}^{2/3}}r^{2}
\]
\[
y_{2}=-\frac{15.2051}{\bar{A}^{2/3}}r^{2}+\frac{12.5116}{\bar{A}^{4/3}}r^{4}
\]
\[
y_{3}=\frac{39.9503}{\bar{A}^{2/3}}r^{2}-\frac{80.3573}{\bar{A}^{4/3}}r^{4}+\frac{37.1429}{\bar{A}^{2}}r^{6}
\]
\[
y_{4}=-\frac{82.4315}{\bar{A}^{2/3}}r^{2}+\frac{293.927}{\bar{A}^{4/3}}r^{4}-\frac{313.522}{\bar{A}^{2}}r^{6}+\frac{103.367}{\bar{A}^{8/3}}r^{8},
\]
where $\bar{A}$ is taken as the average of the mass numbers of the
two isotopes. The sum of the expansion terms has been re-arranged
but ${\displaystyle {\textstyle \sum}_{N=1}^{4}}F_{N}\delta\left\langle r^{2N}\right\rangle ={\textstyle {\displaystyle {\textstyle \sum}_{N=1}^{4}}}c_{N}\delta\left\langle y_{N}\right\rangle $
must still hold. The $c_{N}$ coefficients can be found by equating
same order terms in the above equation. Hence, the new coefficients
are:

\begin{align*}
c_{1}= & 0.288554\bar{A}^{2/3}F_{1}+0.350673\bar{A}^{4/3}F_{1}+\\
+ & 0.448303\bar{A}^{2}F_{3}+0.592709\bar{A}^{8/3}F_{4}\\
c_{2}= & 0.0799258\bar{A}^{4/3}F_{2}+0.172916\bar{A}^{2}F_{3}+0.2972\bar{A}^{8/3}F_{4}\\
c_{3}= & 0.026923\bar{A}^{2}F_{3}+0.08166\bar{A}^{8/3}F_{4}\\
c_{4}= & 0.00967424\bar{A}^{8/3}F_{4}.
\end{align*}
Now, the RFS is given by the summation 
\[
{\displaystyle \sum_{N=1}^{4}}c_{N}\delta\left\langle y_{N}\right\rangle 
\]
and the matching percentage to the final field shift after each term
has been added differs from the one when the original summation is
used. 

As seen in Fig.~9, the orthogonal expansion converges substantially
faster than the original summation. In fact, only the $\delta\left\langle r^{2}\right\rangle $
and $\delta\left\langle r^{4}\right\rangle $ moments need to be considered
as long as the sum is re-arranged. Thus, for a pair of isotopes $A,A'$
and a transition $k$, the RFS is to a very good approximation expressed
as: 
\[
\delta\nu_{k,RFS}^{A,A'}\approx c_{k,1}\delta\left\langle y_{1}\right\rangle +c_{k,2}\delta\left\langle y_{2}\right\rangle .
\]
In case the isotope shifts are known for two transitions, a system
of two equations can be formed, and the $c_{k,1}$ and $c_{k,2}$
constants can be evaluated using the expressions above. They depend
on the line field shift factors $F_{k,N}$ that are different for
each transition and which are calculated for the reference isotope
$A$. Therefore, for two transitions, the problem takes the form of
a matrix equation: 
\[
\left[\begin{array}{c}
\delta\nu_{1,RFS}^{A,A'}\\
\delta\nu_{2,RFS}^{A,A'}
\end{array}\right]\approx\left[\begin{array}{cc}
c_{1,1} & c_{1,2}\\
c_{2,1} & c_{2,2}
\end{array}\right]\left[\begin{array}{c}
\delta\left\langle y_{1}\right\rangle \\
\delta\left\langle y_{2}\right\rangle 
\end{array}\right].
\]
The unknown $y_{1}$ and $y_{2}$ can thus be solved according to:
\[
\left[\begin{array}{c}
\delta\left\langle y_{1}\right\rangle \\
\delta\left\langle y_{2}\right\rangle 
\end{array}\right]\approx C^{-1}\left[\begin{array}{c}
\delta\nu_{1,RFS}^{A,A'}\\
\delta\nu_{2,RFS}^{A,A'}
\end{array}\right],
\]
where $C^{-1}$ is the inverse matrix of $\left[\begin{array}{cc}
c_{1,1} & c_{1,2}\\
c_{2,1} & c_{2,2}
\end{array}\right]$. The $\delta\left\langle r^{2}\right\rangle $ and $\delta\left\langle r^{4}\right\rangle $
moments are finally extracted by solving the equations: 
\[
\left[\begin{array}{c}
\delta\left\langle y_{1}\right\rangle \\
\delta\left\langle y_{2}\right\rangle 
\end{array}\right]=\left[\begin{array}{cc}
3.46556/\bar{A}^{2/3} & 0\\
-15.2051/\bar{A}^{2/3} & 12.5116/\bar{A}^{4/3}
\end{array}\right]\left[\begin{array}{c}
\delta\left\langle r^{2}\right\rangle \\
\delta\left\langle r^{4}\right\rangle 
\end{array}\right].
\]
This can be compared with the original summation, where if the approximate
relation

\[
\delta\nu_{k,RFS}^{A,A'}\approx F_{k,1}\delta\left\langle r^{2}\right\rangle +F_{k,2}\delta\left\langle r^{4}\right\rangle 
\]
is assumed, the matrix equation to be solved is given by: 
\[
\left[\begin{array}{c}
\delta\nu_{1,RFS}^{A,A'}\\
\delta\nu_{2,RFS}^{A,A'}
\end{array}\right]\approx\left[\begin{array}{cc}
F_{1,1} & F_{1,2}\\
F_{2,1} & F_{2,2}
\end{array}\right]\left[\begin{array}{c}
\delta\left\langle r^{2}\right\rangle \\
\delta\left\langle r^{4}\right\rangle 
\end{array}\right].
\]

\label{app-B} 
\end{document}